\documentclass[10pt]{article}
\usepackage[utf8]{inputenc}

\usepackage{amsmath,amsfonts}
\usepackage{algorithm}
\usepackage{graphicx}
\usepackage[left=2cm,right=2cm,top=2cm]{geometry}
\usepackage{textcomp}
\usepackage{xcolor}
\usepackage{cite}
\usepackage{listings}
\usepackage{float}
\usepackage{subfigure}
\usepackage{url}
\usepackage{multirow}
\usepackage{algpseudocode}
\usepackage{etoolbox}
\usepackage{varwidth}
\usepackage{tikz}
\usetikzlibrary{shapes,arrows}
\usepackage{colortbl}
\usepackage{xcolor}

\usepackage{hyperref}

\usepackage{wrapfig}
\usepackage{bbold}
\usepackage{booktabs,multirow}
\usepackage{adjustbox}
\usepackage{tikz}
\usepackage{mathtools}
\usetikzlibrary{quantikz}

\def\ie{{\it i.e.}\hspace{0.1pc}}

\def\etal{{\it et al.}\hspace{0.1pc}}
\def\etc{{\it etc.}\hspace{0.1pc}}

\newcommand{\dicke}[2]{\ket{\smash{D_{#2}^{#1}}}}
\newcommand{\ghz}[1]{\ket{\smash{G_{#1}}}}
\newcommand{\odd}[1]{\ket{\smash{O_{#1}}}}
\newcommand{\even}[1]{\ket{\smash{E_{#1}}}}
\newcommand{\ps}[2]{\ket{\smash{PS_{#2}^{#1}}}}
\newcommand{\dickefidelity}[2]{\bra{\smash{D_{#2}^{#1}}}\rho\ket{\smash{D_{#2}^{#1}}}}
\newcommand{\ghzfidelity}[1]{\bra{\smash{G_{#1}}}\rho\ket{\smash{G_{#1}}}}
\newcommand{\hw}{\mathrm{wt}}
\newcommand{\bigO}{\mathcal{O}}
\newcommand{\Tr}{\mathrm{Tr}}

\begin{document}

\title{Scalable Experimental Bounds for Entangled Quantum State Fidelities}


\author{Shamminuj Aktar$^{1}$,  Andreas Bärtschi$^{2}$, Abdel-Hameed A. Badawy$^{1}$, Stephan Eidenbenz$^{2}$}

\date{
    \small
    $^1$ Klipsch School of Electrical and Computer Engineering, New Mexico State University\\
    $^2$ CCS-3 Information Sciences, Los Alamos National Laboratory\\
    Email: saktar@nmsu.edu, baertschi@lanl.gov,badawy@nmsu.edu, eidenben@lanl.gov
}
\maketitle

\begin{abstract}
Estimating the state preparation fidelity of highly entangled states on noisy intermediate-scale quantum (NISQ) devices is important for benchmarking and application considerations.  
Unfortunately, exact fidelity measurements quickly become prohibitively expensive, as they scale exponentially as $O(3^N)$ for $N$-qubit states, using full state tomography with measurements in all Pauli bases combinations. 
However, Somma~\etal established that the complexity could be drastically reduced when looking at \emph{fidelity lower bounds} for states that exhibit symmetries, such as Dicke States and GHZ States. These bounds must still be tight enough for larger states to provide reasonable estimations on NISQ devices. 
    
For the first time and more than 15 years after the theoretical introduction, we report meaningful lower bounds for the state preparation fidelity of all Dicke States up to $N=10$ and all GHZ states up to $N=20$ on Quantinuum H1 ion-trap systems using efficient implementations of recently proposed scalable circuits for these states. Our achieved lower bounds match or exceed previously reported exact fidelities on superconducting systems for much smaller states. Furthermore, we provide evidence that for large Dicke States $\dicke{N}{N/2}$, we may resort to a GHZ-based approximate state preparation to achieve better fidelity.
This work provides a path forward to benchmarking entanglement as NISQ devices improve in size and quality.
\end{abstract}


\section{Introduction}
Any advantage that quantum computing may have over classic computing usually relies on the principle of superposition, informally defined as being simultaneously in multiple computational basis states. 
Quantum states that are in a superposition such that individual parts cannot be described independently of the state of others are called entangled states. Superposition and entanglement are essential prerequisites for the successful execution of additional gate operations per the overall structure of the quantum algorithm that leads to a desired quantum end state. Thus, proving that a NISQ device produces entangled states is necessary for any quantum computing success. Such low-level testing of quantum mechanical properties is necessary on quantum devices because environmental noise quickly modifies or partially destroys a pure quantum state (as it would be theoretically prepared); this noise process, usually called decoherence, turns the actual state in a computing device into a so-called mixed state, which is a probabilistic mixture of our target pure quantum state and other pure states. The fidelity of a (mixed) quantum state measures to what extent the intended target (pure) quantum state has been realized. Fidelity is defined as `1' if we measure a correct pure quantum state and close to zero, \ie, $1/2^N$ for a maximally mixed $N$-qubit quantum state that consists purely of noise.

Let us use the state $\frac{1}{\sqrt{2}} (\ket{00} + \ket{11})$, \ie, the Bell state as the prototypical entangled state to illustrate the concept of fidelity. 
From a classical vantage point, it may appear tempting to test whether a NISQ device has produced a Bell state as follows: 
We execute a large number of runs of the generating circuit, each with measurement at the end, and count how often we measure $00$, $01$, $10$, and $11$. 
If we measure $00$ and $11$ about 50 percent of the time each, we declare that the devices actually produce the Bell state. 
However, this logic is flawed because any classical device that returns $00$ and $11$ with probability $0.5$ would have also passed such a test. Such a cheat device would be useless and unable to properly execute additional gate operations that propagate the quantum state and its entanglement correctly to execute an entire quantum algorithm. 
An appropriate way to test entanglement is to measure the distance between the actual state prepared on the device and the targeted Bell state. Here, we use the quantum fidelity measure, formally defined in Section~\ref{sec:related-works}.
Characterizing the prepared state, generally a mixed state and thus described by a density matrix, is called quantum state tomography~\cite{dariano2003quantumtomography,altepeter2004quantum}.
It usually requires repeatedly preparing the state and measuring the outcome in all possible combinations of $X$-, $Y$-, and $Z$-axes. This is called full-state tomography and requires $3^N$ tests for $N$-qubit states~\cite{nielsen2010quantum}. The resulting values are fitted to a density matrix using techniques such as maximum likelihood estimation~\cite{james2004measurement,smolin2012efficient} or Bayesian estimation~\cite{buzek1998reconstruction}, resulting in further challenges to quantify errors~\cite{christandl2012reliable,blumekohout2012robust}.

Showing that entanglement exists on NISQ devices across more than $\sim$6 qubits in such a way is computationally prohibitive due to the exponential number of tests. Nevertheless, it remains vital to demonstrate that the device actually leverages quantum mechanical principles.
%
A solution to this dilemma is to approximate or compute upper and lower bounds on the fidelity measure, requiring fewer runs. Somma~\etal~\cite{somma2006lower} discovered such lower bounds for certain symmetric entangled states based on angular momentum operators, with the additional advantage of a smallish polynomial sampling overhead to account for errors given only by statistics of the involved measurements.

\begin{figure}
\centering
\includegraphics[width=.7\linewidth]{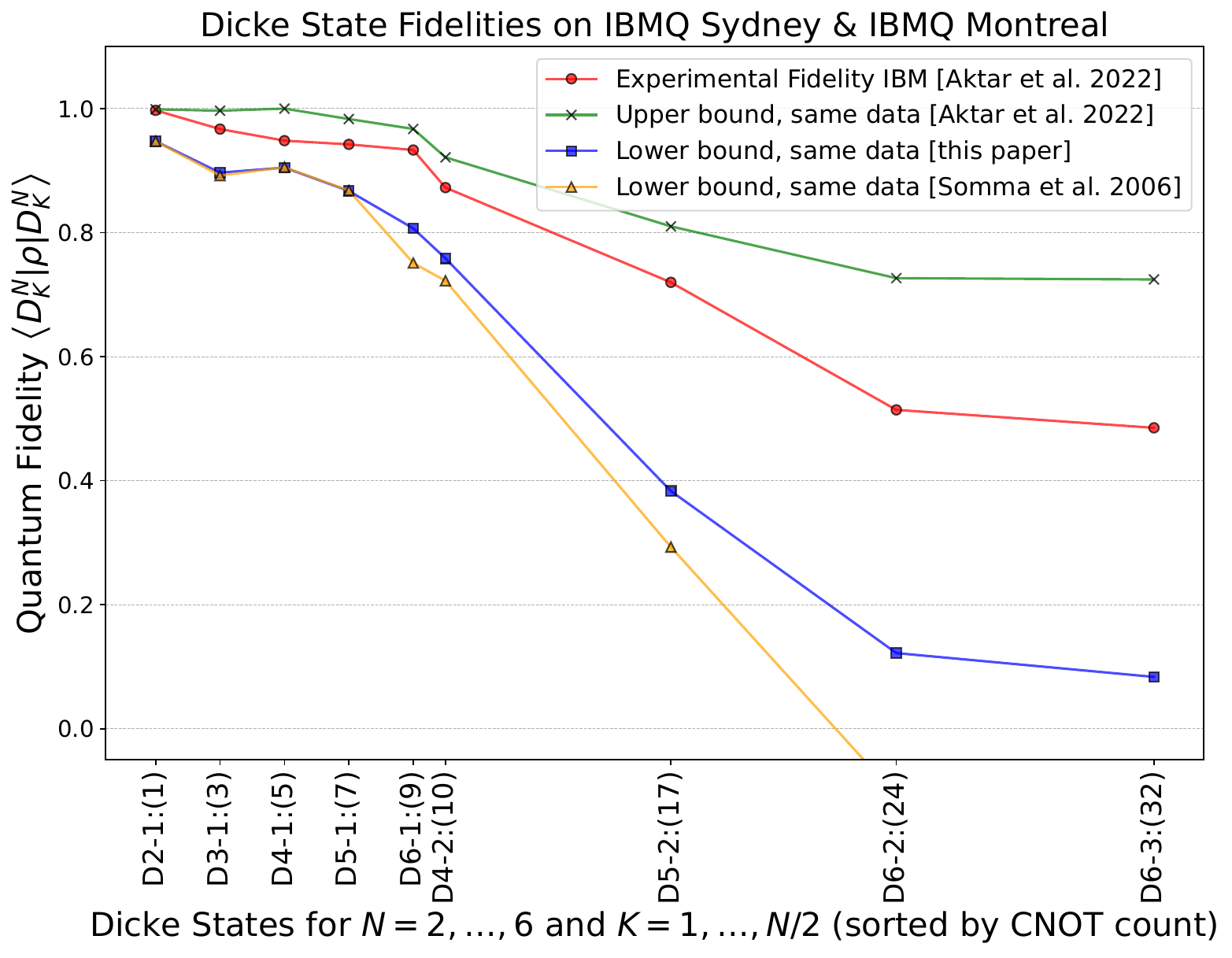}
\caption{Experimental fidelities of Dicke states $\dicke{N}{K}$ (by circuit complexity) 
on IBMQ devices based on full state tomography data~\cite[Aktar~\etal]{aktar2022divide}:
\textit{(red)} Exact fidelity computed from all $3^N$ tomography measurement settings,
\textit{(green)} Upper bound computed only from the $Z$-basis measurement data. 
\textit{(blue)} We compute fidelity lower bounds  from experimental $X$-, $Y$-, and $Z$-basis  measurement setting subsets of their data, 
slightly improving on direct application of existing lower bound techniques~\cite[Somma~\etal]{somma2006lower} $\textit{(orange)}$.%
}
\label{fig:fidelity-comparison}
\end{figure}


In this paper, we adopt and modify these generic bounds for Dicke states and GHZ states -- three well-known classes of entangled states -- as well as approximate Dicke states. Dicke states are equal amplitude superpositions of all computational states of the same Hamming weight (\ie, number of ones). GHZ states are Bell states that are generalized to larger qubit counts. Both Dicke \& GHZ states are studied extensively because of their quantum mechanical properties and their use in various application domains, such as combinatorial constraint optimization.

Through a series of NISQ experiments, we show that the bound from~\cite{somma2006lower} has become useful in practice to provide evidence for entanglement across up to `20' qubits without resorting to exponentially expensive full-state tomography. This is a testament to advances in state preparation algorithms and improvements in NISQ hardware. Figure~\ref{fig:fidelity-comparison} shows that the original bounds were not yet useful for NISQ devices, even as recent as 2022 for Dicke states on as few as `5' qubits (as they give negative values). They become slightly positive with our modifications without providing strong evidence of entanglement; however, as we will show, newer devices (Quantinuum) produce results that are good enough for these bounds to become meaningful lower bounds for fidelity. The original and our modified fidelity bounds require only very few different measurement runs (three measurement settings for Dicke states and two for GHZ states), thus making fidelity-based entanglement verification easy on NISQ devices, even at high qubit counts.
The contributions of the paper are as follows%
\footnote{Parts of our experimental bounds on Dicke and GHZ state fidelity have appeared in an extended abstract at ACM Computing Frontiers 2023~\cite{10.1145/3587135.3592197}.}:
\begin{itemize}
    \item We propose an improved divide-and-conquer based method for efficient Dicke state preparation with $\bigO(N)$ circuit depth and $\bigO(KN)$ gate count with low constant factors. 

    \item We utilize and improve the lower bounds on the quantum fidelity from theory~\cite{somma2006lower} using only three measurement settings for Dicke states and only two for GHZ states.

    \item We demonstrate, using large-scale experiments on Dicke and GHZ state preparation on Quantinuum's H1 ion-trap quantum device, the usefulness of the bounds for the first time on real quantum hardware.

     \item We report meaningful lower bound for all Dicke States up to `10' qubits and all GHZ states up to `20' qubits using efficient circuit implementations.
    
    \item We give state preparation fidelities that match or exceed exact fidelity records. For example, we provide state preparation fidelity lower bounds of (i) 0.46 for the `10' qubit Dicke State $\dicke{10}{5}$ and (ii) 0.73 for the `20' qubit GHZ State $\ghz{20}$. These match or exceed exact fidelity records recently achieved on IBM's superconducting systems for much smaller states $\dicke{6}{3}$~\cite{aktar2022divide}, and $\ghz{5}$~\cite{epfl2019wstate}, respectively. 

    \item Additionally, we show an alternative method for preparing larger Dicke state $\dicke{N}{K}$ approximately, using product states $\ps{N}{K}$ for small $K$ and Even/Odd Hamming weight states $\even{N}$, $\odd{N}$ for large $K$ with even/odd parity. 
    
    \item We show those approximate states can give a good approximation of larger Dicke states by computing lower bounds for product state $\ps{N}{K}$ up to `10' qubits \& Hamming weight $N/2$ and odd/even Hamming weight states $\odd{N}$, $\even{N}$ up to `10' qubits. 
\end{itemize}

The rest of the paper is organized as follows: Section~\ref{sec:related-works} summarizes the related works on fidelity estimation. Section~\ref{sec:methodology} includes our experimental methodology. Sections~\ref{sec:dicke} \&~\ref{sec:ghz} review the state preparation, lower bound estimation, and experimental results for Dicke and GHZ states. Section~\ref{sec:approx-dicke} includes product state and odd/even state preparation as approximate Dicke state, lower bound computation, and their experimental results.
Section~\ref{sec:conclusion} concludes the paper.

\section{Related Work}
\label{sec:related-works}
We use the \emph{quantum fidelity} $\mathcal{F}$ as a similarity measure between a prepared mixed state $\rho$, expressed as a density matrix,
and a target pure quantum state $\rho_{\psi} = \ket{\psi}\bra{\psi}$~\cite{jozsa1994fidelity}:
\begin{align}
	\mathcal{F}(\rho_{\psi},\rho) = \left[ \Tr\sqrt{\sqrt{\rho}\rho_\psi\sqrt{\rho}}\right]^2 = \Tr(\rho_{\psi}\rho) = \bra{\psi}\rho\ket{\psi}
\end{align}
where $\sqrt{\rho}$ denotes the unique positive semi-definite square root of $\rho$ such that $\sqrt{\rho}\sqrt{\rho} = \sqrt{\rho}^{\dagger}\sqrt{\rho}=\rho$, 
where the first equality would also hold for a mixed state $\rho_{\psi}$.

In this paper, we consider pure target states, notably Dicke states $\dicke{N}{K}$, and GHZ states $\ghz{N}$. 
We remark that quantum fidelity is sometimes also defined as $\mathcal{F'} = \sqrt{\mathcal{F}}$~\cite{nielsen2010quantum,somma2006lower}. 
In a straightforward way, the fidelity of a prepared $N$-qubit state can be computed by sampling $\rho$ in $3^N$ different Pauli basis 
($N$-fold tensor products of Pauli operators $\sigma_x$,$\sigma_y$ and $\sigma_z$) to reconstruct its density matrix; hence, we can compute the fidelity with the target state.

Suppose we want to \emph{upper bound} the quantum fidelity. In that case, we can consider the \emph{measurement success probability}, \ie, the overall probability of sampling non-zero amplitude states when measuring in the computational $Z$-basis: 
\begin{align}
	\mathit{MSP}(\rho_{\psi},\rho) := \sum\nolimits_{x \in \left\{0,1\right\}^n,\ \bra{x}\rho_{\psi}\ket{x}\neq 0} \bra{x}\rho\ket{x}
\end{align}
Another upper bound measure is \emph{Hellinger Fidelity}, which quantifies the similarity between the probability distributions of the pure target state and the measurement probabilities in the computational $Z$-basis:
\begin{align}
\mathit{H}(\rho_{\psi}, \rho) := \left[ \sum_{x \in \left\{0,1\right\}^N} \sqrt{\bra{x}\rho_{\psi}\ket{x} \cdot \bra{x}\rho\ket{x}} \right]^2
\label{h-eq}
\end{align}
Using a double application of the Cauchy-Schwarz inequality, one gets $\mathcal{F}(\rho_\psi,\rho) \leq \mathit{H}(\rho_\psi,\rho) \leq \mathit{MSP}(\rho_\psi,\rho)$~\cite{aktar2022divide}.

\newcommand{\ryangle}[3]{\scriptstyle\smash{R_y(#3#2\cos^{-1}(\sqrt{#1}))}}
\newcommand{\rygate}[3]{\gate{\ryangle{#1}{#2}{#3}}}	   

Several works have shown that it is possible to estimate a \emph{lower bound} on the quantum fidelity using only a few measurement settings, avoiding full-state tomography. Earlier works~\cite{somma2006lower, guhne2007toolbox} focused on certain states with unique symmetry that require only a small random subset of Pauli operators to estimate fidelity.
Somma~\etal~\cite{somma2006lower} proposed expressions for estimating quantum fidelity for highly symmetric classes of multi-qubit state preparation, \ie, rotational invariant states, stabilizer states, and generalized coherent states. To estimate the lower fidelity bound, they only used the measurements related to the symmetry operators of the density matrix.
Guhne~\etal~\cite{guhne2007toolbox} derived fidelity estimation for $GHZ$ and $W$ states using $2N-1$ measurements. Flammia~\etal~\cite{flammia2011direct} had a generalized approach where they showed estimation techniques for all possible state preparation by measuring fewer Pauli operators close to the desired state and of greater importance. Mentioning the scalability issue in generalized $N$ qubit state fidelity estimation, Elben~\etal~\cite{elben2022randomized} showed how randomized measurement could enable fidelity estimation by comparing two states. Zhang~\etal~\cite{zhang2021direct} proposed a machine learning (ML) approach for direct fidelity estimation as a classification problem using a constant number of expectations. Recently, a \emph{classical shadow} approach has been introduced to enhance tomography efficiency, with potential benefits for fidelity estimation for certain quantum states~\cite{huang2020predicting,struchalin2021experimental}.

Additionally, quantum state verification (also known as the non-tomographic method) uses advanced statistical approaches to verify whether the output of some device is the target state. Previous works on quantum state verification techniques validate  specific state preparation such as Dicke states~\cite{liu2019efficient}, GHZ states~\cite{li2020optimal}, Hypergraph states~\cite{zhu2019efficienthyp}, Pure states~\cite{li2019efficient,yu2019optimal,zhu2019efficient,wang2019optimal}, \etc 
There are also some generalized approaches for state verification~\cite{zhu2019general, zhu2019optimal, pallister2018optimal,zhang2020experimental, jiang2020towards,  tokunaga2006fidelity}. Furthermore, Yu~\etal~\cite{yu2022statistical} utilizes advanced statistical methods for resource-efficient quantum state verification.

\section{Methodology}
\label{sec:methodology}
In this work, we run the experiments on both the hardware and emulator devices of Quantinuum H-systems~\cite{quantinuum} using the Quantinuum Python API. Running jobs on the Quantinuum backend requires H-System Quantum Credits (HQC) that are linear with both the number of shots per circuit and the number of CNOTs in the circuit. The total cost generally scales with the product of the number of shots and CNOT counts. Our initial goal was to determine the number of experiments and the number of shots for different state preparations permissible within the HQCs allocated to us. As the Quantinuum machines are available through a calendar month and HQCs are allocated monthly, we target to complete experiments throughout several months with available hardware credits. 

First, we run the experiments on the Quantinuum H1-2E emulator to estimate the confidence interval sizes we can expect from the quantum hardware. To see the effect of the number of samples/circuit executions, we compute the cumulative bounds for fidelity and measurement success probability. Similarly, we compute a two-sided 68\% confidence interval below and above the estimator for the mean based on Student's t-distribution to check the bounds' stability with an increasing number of shots. Analyzing the results of emulator experiments, we get an approximation of a meaningful number of experiments that can be performed on quantum hardware for various states, improving on the (polynomial) number of samples derived from theoretical considerations (as elaborated on later). We limit Dicke state preparation up to $N = 10$ as Dicke States starting at $N = 11$ become prohibitively expensive computationally. For GHZ states, we go up to $N=20$ as we run the experiment on a 20-qubit H1-1 device. Additionally, we prepare 
approximate Dicke states up to $N=10$ qubits using product states for $K = 1$ to $N/2$ and odd/even states for $N = 2$ to $10$ and $N = 4$ to $10$ respectively.

We observe roughly equal confidence interval widths for the bounds across the experiments, when we use a constant number of $2\times 200$ samples for all GHZ states, and $3\times\max(150, 4\tbinom{N}{K})$ samples for Dicke states $\dicke{N}{K}$. Next, depending on their availability, we execute the experiments of different state preparations on Quantinuum hardware backends. Dicke state circuits are executed in the `12' qubit H1-2 processor, while GHZ state circuits are executed in the `20' qubit H1-1 processor. For all experiments, we generate the circuits using IBM Qiskit~\cite{ibm-qiskit} and directly submit the QASM~\cite{cross2017open} code using the Python API. To execute the circuits in the same time span, we submit the jobs in batch mode; thus, we wait for the queue time once. However, the Quantinuum queue only allows 500 HQC execution in a single batch. Thus, we execute sub-experiments over a couple of months. 
As a mixed state, which can mathematically be considered a probability distribution over states, 
the definition of the prepared state's fidelity fully captures drifts in the device over different time slots used to prepare the state.%
\footnote{Dicke states experiments were run in this way: $\dicke{2}{1}$ to $\dicke{7}{2}$ on May 24th, $\dicke{8}{1}$ to $\dicke{9}{2}$ on June 6th, $\dicke{9}{3}$ to $\dicke{10}{3}$ on June 10th, $\dicke{10}{4}$ on June 17th and $\dicke{10}{5}$ on July 25th of 2022. Logarithmic and linear depth GHZ circuits were executed on August 19th, 2022. Approximate Dicke states (odd/even states and product states) were executed on July 12th, 2022.}

\section{Fidelity Lower Bounds for Dicke States}
\label{sec:dicke}
We provide experimental lower bound estimations on the state preparation fidelity of Dicke states $\dicke{N}{K}$ (up to $N=10$). We first discuss the circuit preparation strategies, then provide lower bound expressions for estimating quantum fidelity, and present experimental results for fidelity estimation in Quantinuum's H1-2 device.

\subsection{State Preparation}
\label{dicke-prep}
A Dicke state $\ket{D_K^N}$ is the equal weight superposition of all $\binom{N}{K}$ $N$-qubit basis states $x$ with $K$ Ones and $N-K$ Zeroes, \ie, with Hamming weight $\hw(x) = K$:
\begin{align}
    \label{eq:dicke}
    \dicke{N}{K} = \tbinom{N}{K}^{-1/2} \sum\nolimits_{x \in \left\{ 0,1 \right\}^N,\ \hw(x)=K}{\ket{x}}.
\end{align}
We optimize and generalize existing work on Dicke state preparation by defining Dicke state unitaries $U_{k, K}^N$, which prepare Dicke states $\dicke{N}{\ell}$ 
upon any input of a unary encoded Hamming weight $k \leq \ell \leq K$:
\begin{align}
	U_{k,K}^N\colon \ket{0^{N-\ell}1^{\ell}} \mapsto \dicke{N}{\ell} \qquad \forall k\leq \ell \leq K.
\end{align}
Dicke state unitaries were first defined for $k:=0$~\cite{baertschi2019deterministic},
and there exists a linear-depth circuit construction for $k:=0,\ K:=N$ with $5\binom{N-1}{2} + 2(N-1)$ two-qubit CNOT gates 
between neighboring qubits in a 1D Linear Nearest Neighbor (LNN) connectivity only~\cite{aktar2022divide}. 
We start from this construction and further reduce the CNOT count by $3\binom{N-K-1}{2}$ and $5\binom{k}{2}$ using ideas for upper~\cite{baertschi2019deterministic} 
and lower~\cite{mukherjee2020preparing} bounds on the Hamming weight $\ell$, respectively. 
This effectively generalizes and improves all former constructions by constant factors while preserving LNN connectivity.

To prepare a fixed-$K$ Dicke state $\dicke{N}{K}$, we further parallelize our construction using the  divide-and-conquer  strategy in ~\cite{baertschi2022shortdepth,aktar2022divide}: 
The main idea is to split the Hamming weight $K$ across qubit sets of size $N_1 = \lfloor \tfrac{N}{2} \rfloor$ and $N_2 = \lceil \tfrac{N}{2} \rceil$ using $2K-1$ CNOT gates
on an all-to-all connectivity before applying the Dicke state unitaries $U_{0,K}^{N1}$ and $U_{0,K}^{N2}$ in parallel.
Additional gains can be made by implementing these unitaries with layers of bit-flipping Pauli-$X$ gates, $U_{0, K}^{N_i} = X^{\otimes N_i} \cdot U_{N_i-K, N_i}^{N_i} \cdot X^{\otimes N_i}$,
resulting in the circuit structure shown in Figure~\ref{fig:dicke}, 
which prepares $\dicke{9}{3}$ with the following steps: 
\begin{align*}
	\ket{0^5}\ket{0^4}	& \xmapsto{(1)} \tbinom{9}{3}^{-1/2} \sum\nolimits_{\ell=0}^3	\sqrt{\tbinom{5}{3-\ell}\tbinom{4}{\ell}}\ket{00000}\ket{1^\ell0^{3-\ell}0}		\\
				& \xmapsto{(2)} \tbinom{9}{3}^{-1/2} \sum			\sqrt{\tbinom{5}{3-\ell}\tbinom{4}{\ell}}\ket{0^{3-\ell}1^\ell11}\ket{1^\ell0^{3-\ell}0}	\\
				& \xmapsto{(3)} \tbinom{9}{3}^{-1/2} \sum			\sqrt{\tbinom{5}{3-\ell}\tbinom{4}{\ell}}\ket{0^{3-\ell}1^\ell11}\ket{0^\ell1^{3-\ell}1}	\\
				& \xmapsto{(4)} \tbinom{9}{3}^{-1/2} \sum			\sqrt{\tbinom{5}{3-\ell}\tbinom{4}{\ell}}\dicke{5}{2+\ell}\dicke{4}{4-\ell}		\\
				& \xmapsto{(5)} \tbinom{9}{3}^{-1/2} \sum			\sqrt{\tbinom{5}{3-\ell}\tbinom{4}{\ell}}\dicke{5}{3-\ell}\dicke{4}{\ell} =\dicke{9}{3}.
\end{align*}
Note that to prepare the Dicke state $\dicke{9}{6}$, we can simply omit the last layer of Pauli-$X$ gates. 
Thus, in our experiments we only prepare Dicke states $\dicke{N}{K}$ up to $K\leq \tfrac{N}{2}$, for which our two-qubit gates amount to $5K(N-K)-3(N+K)+5$ CNOTs, giving linear depth $\bigO(N)$ and total gate count $\bigO(KN)$.

\begin{figure*}[t!]
	\centering
	\begin{adjustbox}{width=0.99\linewidth}
	   \begin{quantikz}[row sep={21pt,between origins},execute at end picture={}]
		\lstick{$q_0\colon\ket{0}$}	& \qw						& \qw		& \qw			& \qw		& \qw\slice{(1)}		& \qw		& \qw		& \qw		& \qw\slice{(2)}	& \gate{X}\slice{(3)}	& \gate[4]{U_{1,4}^{4}}\slice{(4)}	& \gate{X}\slice{(5)}	        & \qw\rstick[9]{\rotatebox{90}{$\dicke{9}{3}$}}\\
		\lstick{$q_1\colon\ket{0}$}	& \rygate{\phantom{3}4/34}{\phantom{2}}{} 	& \qw		& \qw			& \targ{}	& \rygate{\phantom{3}4/34}{}{-}	& \qw		& \qw		& \qw		& \ctrl{7}	        & \gate{X}		& \qw					& \gate{X}		        & \qw\\
	    	\lstick{$q_2\colon\ket{0}$}	& \rygate{34/74}{\phantom{2}}{}			& \targ{}	& \rygate{34/74}{}{-}	& \ctrl{-1}	& \qw				& \qw		& \qw		& \ctrl{5}	& \qw		        & \gate{X}		& \qw					& \gate{X}		        & \qw\\
	    	\lstick{$q_3\colon\ket{0}$}	& \rygate{10/84}{2}{}	   			& \ctrl{-1}	& \qw			& \qw		& \qw				& \qw		& \ctrl{3}	& \qw		& \qw			& \gate{X}		& \qw					& \gate{X}			& \qw\\
		\lstick{$q_4\colon\ket{0}$}	& \qw						& \qw		& \qw			& \qw		& \qw				& \gate{X}	& \qw		& \qw		& \qw		        & \qw			& \gate[5]{U_{2,5}^{5}}			& \gate{X}		        & \qw\\
		\lstick{$q_5\colon\ket{0}$}	& \qw						& \qw		& \qw			& \qw		& \qw				& \gate{X}	& \qw		& \qw		& \qw		        & \qw			& \qw					& \gate{X}		        & \qw\\
		\lstick{$q_6\colon\ket{0}$}	& \qw						& \qw		& \qw			& \qw		& \qw				& \qw		& \targ{}	& \qw		& \qw		        & \qw			& \qw					& \gate{X}		        & \qw\\
		\lstick{$q_7\colon\ket{0}$}	& \qw						& \qw		& \qw			& \qw		& \qw				& \qw		& \qw		& \targ{}	& \qw		        & \qw			& \qw					& \gate{X}		        & \qw\\
		\lstick{$q_8\colon\ket{0}$}	& \qw						& \qw		& \qw			& \qw		& \qw				& \qw		& \qw		& \qw		& \targ{}	        & \qw			& \qw					& \gate{X}		        & \qw
	    \end{quantikz}
	\end{adjustbox}
        \caption{%
		Divide-and-conquer Dicke state $\dicke{9}{3}$ preparation allowing parallelizable Dicke state unitaries on $4$ and $5$ qubits:
		\textit{(1)} We prepare a correctly weighted superposition of input Hamming weights $0\leq\ell\leq 3$ on the first register. 
		The used $R_y$-rotations contain arguments with numerators (denominators) derived from (suffix-sums) of terms of the form $\tbinom{5}{3-\ell}\tbinom{4}{\ell}$, 
		the number of distributions of $\ell$ \& $3-\ell$ Ones across 4 \& 5 qubits. 
		\textit{(2\&3)} The first register is correctly entangled with the second register, and bit-flipping $X$-gates are applied to reduce the number 
		of CNOTs in the following. 
		\textit{(4\&5)} Parallel Dicke state unitaries $U_{1,4}^4$ and $U_{2,5}^5$ prepare the Dicke state $\dicke{9}{6}$, followed by bit-flipping $X$-gates to get $\dicke{9}{3}$.
	}
	\label{fig:dicke}
\end{figure*}

\subsection{Lower bounds for fidelity estimation}
\label{lb-dicke}
Following~\cite{somma2006lower}, we observe that
Dicke States are rotationally invariant and thus are the unique simultaneous eigenstates of the (squared) total angular momentum $J^2 = J_x^2 + J_y^2 + J_z^2$ and its $z$-component $J_z$, where $J_{\tau} = \sigma_{\tau}^1 + \sigma_{\tau}^2 + ... +\sigma_{\tau}^N$ for Pauli operators $\sigma_\tau^i$, $\tau=x,y,z$, acting on qubit $i$. 
This means we can write the Dicke state $\dicke{N}{K}$ as $\ket{j,j_z}$ in terms of its quantum numbers $j=N$ and $j_z=N-2K$ with:
\begin{align}
    \label{eq:J2}
    J^2\ket{j,j_z}	&= j(j+2)\ket{j,j_z}	\\
    \label{eq:Jz}
    J_z\ket{j,j_z}	&= j_z\ket{j,j_z}.
\end{align}
In general, the quantum numbers $j$ and $j_z$ satisfy $0\leq j \leq N$ and $|j_z| \leq j$. The difference between any two eigenvalues is at least 2
(because we use Pauli operators instead of spin-1/2 operators). 

Now, we can write a target state $\rho_{j,j_z} := \ket{j,j_z}\bra{j,j_z}$ in terms of the operators $J^2$ and $J_z$:
\begin{align}
	\label{rho}
	\rho_{j,j_z} = 
	\prod_{\substack{j_z' \neq j_z \\ -j \leq j_z' \leq j}} \frac{J_z - j_z'}{j_z-j_z'} 
	\prod_{\substack{j' \neq j \\ 0 \leq j' \leq N}} \frac{J^2 - j'(j'+2)}{j(j+2)-j'(j'+2)}.
\end{align}
We could compute the fidelity $\mathcal{F}(\rho_{j,j_z},\rho)$ by measuring every Pauli-String appearing when we expand the product, \ie, all correlations between the operators $J^2$ and $J_z$ in Equation~\eqref{rho}. 
This process requires measuring an exponentially large number of observables but less than $3N$ measurements for performing full-state tomography. 

Instead, we turn towards the construction of operators that allow us to compute a lower bound on the fidelity. 
We will define operators $\mathcal{S}_{J_z}$ and $\mathcal{S}_{J^2}$ inspired by the density operator of Equation~\eqref{rho} such that they satisfy
$[\mathcal{S}_{J_z} + \mathcal{S}_{J^2}] \ket{j',j_z'}= e_{j',j_z'} \ket{j',j_z'}$ with $e_{j',j_z'} = 1$ for the correct Dicke state and $e_{j',j_z'} \leq 0$ for $(j',j_z')\neq (j,j_z)$. 
We define the operator $\mathcal{S}_{J_z}$ as the left-hand product in Equation~\eqref{rho},
\begin{align}
    \label{sjz}
    \mathcal{S}_{J_z} := \prod\nolimits_{j_z' \neq j_z} [ (J_z - j_z') / (j_z - j_z') ],
\end{align}
for which $\ket{j',j_z'}$ has eigenvalue 1 if $j_z' = j_z$ and 0 otherwise.%
\footnote{%
We note that in~\cite{somma2006lower}, the operator is defined as $\mathcal{S}_{J_z} = -\frac{1}{4}(J_z - j_z))^2 + 1$, 
which for states $\ket{j',j_z'\neq j_z}$ leads to eigenvalues $\leq$ 0 instead of a strict equality $=$ 0. 
This results in less tight bounds; see Figure~\ref{fig:fidelity-comparison}.
}
Expanding the product results in only Pauli-Strings consisting of identities and $\sigma_z$ operators. These commute pairwise, 
hence, we can measure $\mathcal{S}_{J_z}$ in the computational basis and thus in a single measurement setting. 
In fact, we get the measurement success probability $\Tr\left[\mathcal{S}_{J_z} \cdot \rho\right] = \mathit{MSP}(\rho_{j,j_z},\rho)$. 

The same is not true if we expanded the right-hand product of Equation~\eqref{rho}: We would get a large number of non-commuting Pauli-Strings. 
Instead we use the following definition for $\mathcal{S}_{J^2}$~\cite{somma2006lower}, which is based on the fact that Dicke states 
have maximum angular momentum, and all other eigenvalues for $J^2$ deviate by at least $N(N+2)-(N-2)N = 4N$:
\begin{figure*}[t!]
\centering
\includegraphics[width=0.99\textwidth]{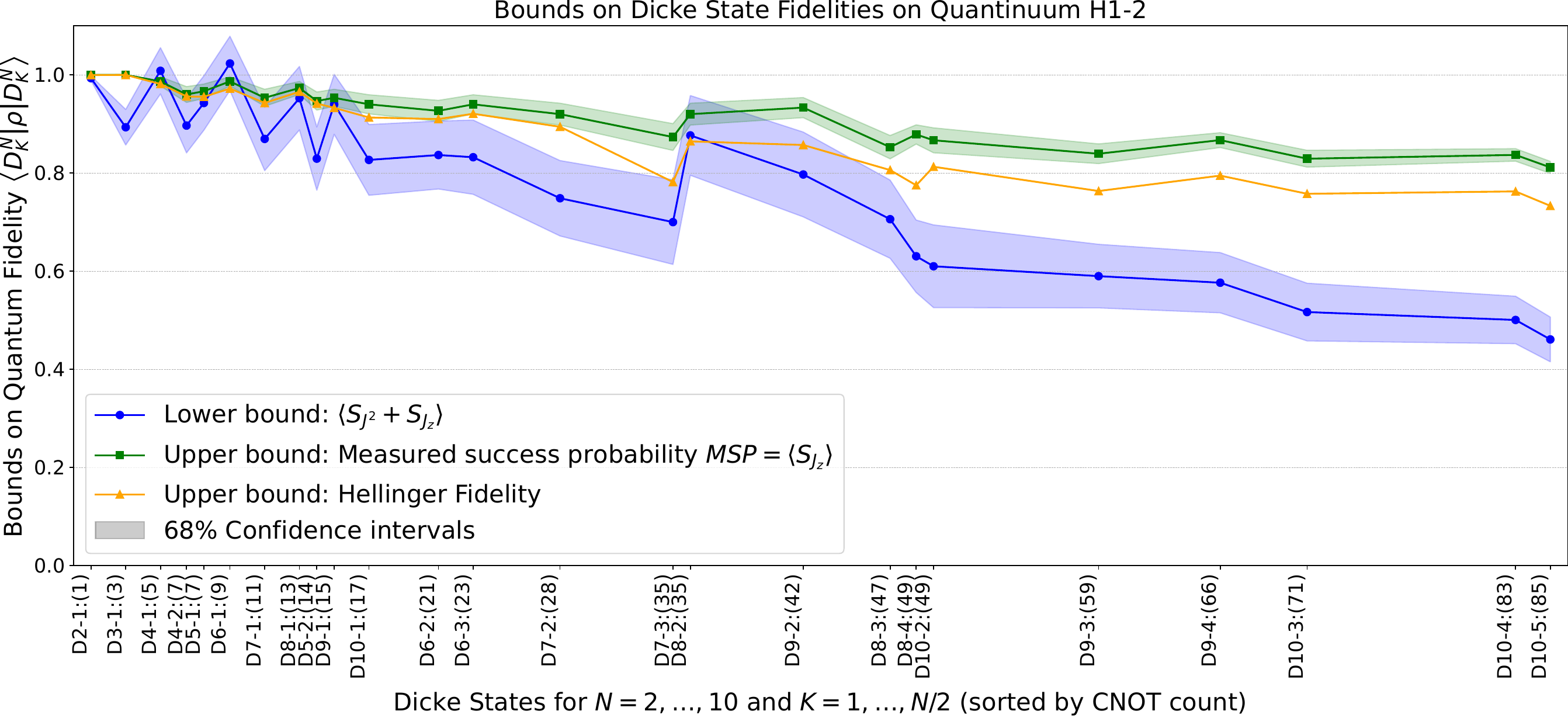}\\[2ex]
\includegraphics[width=0.99\textwidth]{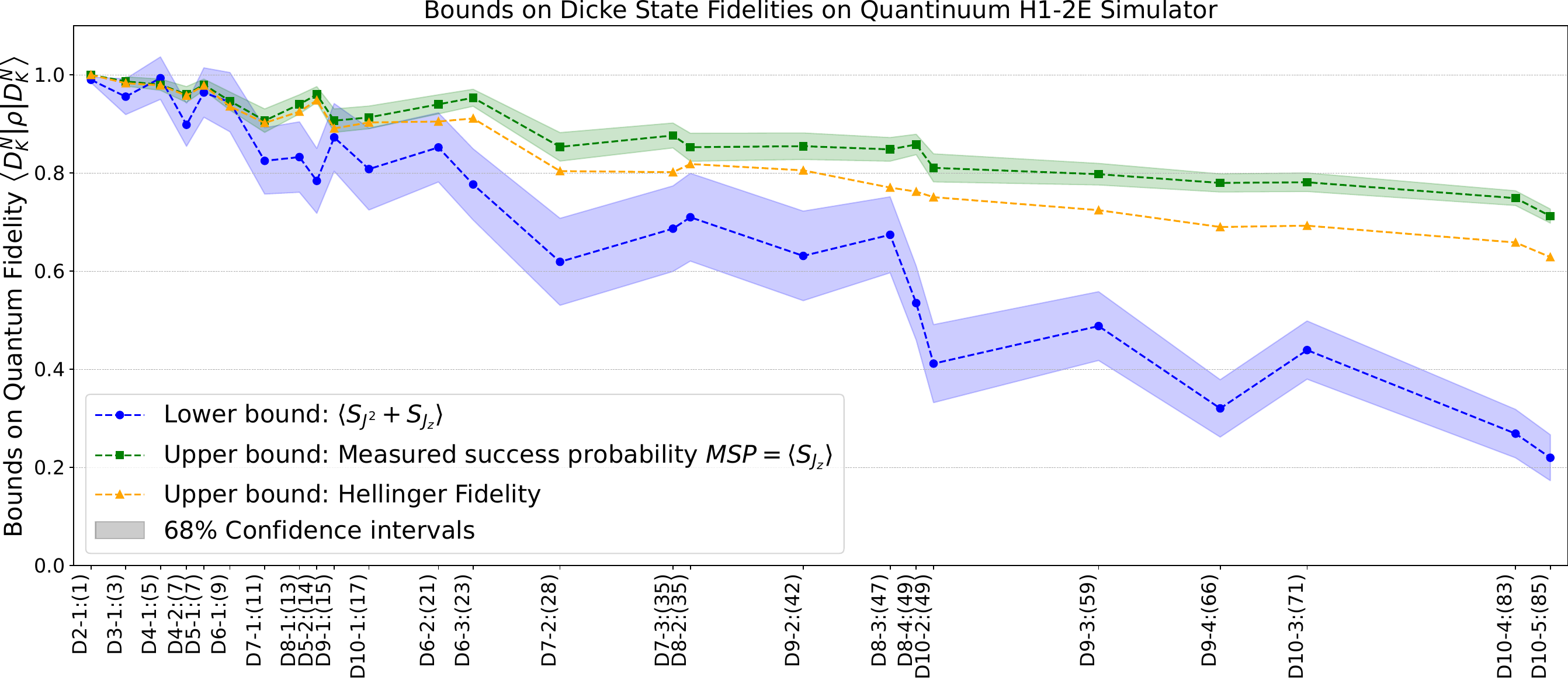}
\caption{Upper and lower bounds on the quantum fidelity $\dickefidelity{N}{K}$ for Dicke States $\dicke{N}{K}$ on the Quantinuum H1-2 Processor~\textit{(top)} and its H1-2E Simulator~\textit{(bottom)}, including 68\% confidence intervals.
Dicke states are sorted along the $x$-axis according to the number of CNOTs in their preparation circuits. Compared to the simulator, the Quantinuum H1-2 Processor demonstrates better upper and lower bounds on the Dicke state preparation fidelity.
The bump in fidelity on Quantinuum H1-2 for Dicke states $\dicke{8}{1}$--$\dicke{8}{4}$ and $\dicke{9}{1}$--$\dicke{9}{2}$ may be due to these experiments taking place right after a recalibration downtime of the QPU.
}
\label{fig:dicke-all}
\vspace*{-1ex}
\end{figure*}

\begin{align}
    \label{eq:sj2}
    \mathcal{S}_{J^2} = \tfrac{1}{4N}(J^2 - N(N+2)), 
\end{align}
for which $\ket{j',j_z'}$ has eigenvalue 0 if $j' = j =  N$ and integer eigenvalues $\leq -1$ otherwise.
Similar to $\mathcal{S}_{J_z}$, we can measure the operator $\mathcal{S}_{J^2}$ by computing $\Tr \left[ J_\tau^2\cdot \rho \right]$ through measurement in the $\tau$-basis, \ie, 
measuring in the three $X$-basis, $Y$-basis and $Z$-basis settings is enough.

If the sum of the operators $\mathcal{S}_{J_z}$ and $\mathcal{S}_{J^2}$ are applied to a target Dicke state $\rho = \ket{j',j_z'}$, 
we get eigenvalue $1$ for the correct Dicke state and eigenvalues $\leq 0$ for all other rotational states. 
In general, if $\ket{\phi} = \sum_{j',j_z'} c_{j',j_z'} \ket{j',j_z'}$ is a pure state, then we get
\begin{align}
    \label{eq:less}
    \bra{\phi}\mathcal{S}_{J_z} + \mathcal{S}_{J^2}\ket{\phi} = \sum\nolimits_{j',j_z'} (e_{j',j_z'})|c_{j',j_z'}^2| \leq |c_{j,j_z}|^2
\end{align} 
Here $|c_{j,j_z}|^2$ is the probability of projecting $\ket{\phi}$ onto the state $\ket{j=N,j_z = N-2K}$. 
Our experimentally prepared mixed state $\rho$ is a convex combination of pure states, yielding our lower bound
\begin{align}
    \label{eq:dicke-fidelity}
	 \langle \mathcal{S}_{J_z}+\mathcal{S}_{J^2}\rangle_\rho:= \Tr[(\mathcal{S}_{J_z}+\mathcal{S}_{J^2})\cdot\rho] \leq \mathcal{F}(\rho_{j,j_z},\rho) 
\end{align}

\subsection{Experimental Results}
\label{sec:dicke-exp}
We construct Dicke state \ket{D_K^N} circuits using the preparation scheme described in Section~\ref{dicke-prep}. 
To obtain the bounds on fidelity, first, we generate the untranspiled QASM circuits for Dicke State $\dicke{N}{K}$ where $2 \leq N \leq 10$ and $1 \leq K \leq N/2$. Next, we compute the bounds on the quantum fidelity using only three measurements ($X$, $Y$, and $Z$ basis measurement). 
We use the formulas of Section~\ref{lb-dicke} to estimate the lower bound ${\langle \mathcal{S}_{J_z} + \mathcal{S}_{J^2} \rangle}_{\rho}$ on quantum fidelity $\dickefidelity{N}{K}$. We also show the probability of finding expected basis states among the total number of circuit executions (informally called measured success probability). The definition of $S_{J_z}$ in \ref{sjz} shows that it is the measured success probability of the prepared Dicke state. Additionally, we show the Hellinger fidelity as an upper bound to the state preparation fidelity for the prepared states.
\subsubsection{Overall Results}

Figure~\ref{fig:dicke-all} shows the bounds for Dicke State $\dicke{N}{K}$  on Quantinuum H1-2E simulator and H1-2 processor where the Dicke states from left to right are sorted according to the number of required CNOT gates in the untranspiled circuit. In the $X$ axis, Dicke states $\dicke{N}{K}$ are represented as  $D-N-K:(C)$ where $N$ is the number of qubits, $K$ is Hamming weight and $C$ is the number of CNOT counts. Each plot shows the lower bound ${\langle \mathcal{S}_{J_z} + \mathcal{S}_{J^2} \rangle}_{\rho}$, measured success probability along with a two-sided 68\% confidence interval below and above the mean of the distribution, and Hellinger fidelity. We observe statistical noise in the lower bound estimation for the smaller Dicke states (with few CNOTs). We also find that the fidelity lower bound decreases with the increasing number of CNOT gates on both the simulator and real hardware. Also, we observe that the bounds on Dicke state preparation fidelity are better in hardware than in the simulator. For our largest Dicke state $\dicke{10}{5}$, the simulator lower bound estimation on fidelity is 0.22 while the hardware estimation is 0.46. Quantinuum specification mentions that although the simulators provide a high-fidelity representation of the hardware device's output, some variances between the results may occur as the simulation's noise models cannot fully capture the actual hardware behavior.
Additionally, we observe that both the upper bounds (measured success probability and Hellinger fidelity) decrease with increasing CNOT count, although the decrease is much flatter than the lower bound. Notably, the Hellinger fidelity provides a much sharper bound than the measured success probability. Initially, the Hellinger fidelity is very close to the measured success probability. Still, as the CNOT count increases, the gap between the Hellinger fidelity and the measured success probability starts to widen. Furthermore, the gap between the lower and the upper bounds increases for higher Dicke states. Overall, the experimental results for the bounds on both the simulator and hardware follow the theoretical relationships among the bounds mentioned in Section~\ref{sec:related-works}.

The sudden inconsistency of lower bound estimation from $\dicke{8}{2}$ on hardware could be because the experiments before and after $\dicke{8}{2}$ were executed in a different time span. Moreover, the experiments after $\dicke{7}{3}$ were run just after the H1-2 machine became online after calibration.

\subsubsection{Number of Shots}
\emph{Theoretical.}
To determine a sufficient number of shots $N_s$ per projective measurement, we first derive an upper bound in terms of a $\gamma=68$\%-confidence interval of width at most $2 \varepsilon=0.2$. We can write the fidelity lower bound from Equation~\eqref{eq:dicke-fidelity} as
\begin{align}
    \mathcal{F}(\rho_{j,j_z},\rho) \geq \langle \mathcal{S} \rangle_{\rho} := \langle \mathcal{S}_{J_z}+\mathcal{S}_{J^2}\rangle_{\rho}
    &= \left( \langle \mathcal{S}_{J_z} \rangle_{\rho} +\ \langle \tfrac{J_z^2}{4N} \rangle_{\rho} \right) \quad\ +\ \langle \tfrac{J_x^2}{4N} \rangle_{\rho} \qquad +\ \langle \tfrac{J_y^2}{4N} \rangle_{\rho} \qquad - \tfrac{N+2}{4}    \label{eq:dicke-fidelity-commuting}    \\
    &= \mathit{MSP}(\rho) + \sum_{i<j} \langle \tfrac{\sigma_z^i \sigma_z^j}{2N} \rangle_{\rho} + \sum_{i<j} \langle \tfrac{\sigma_x^i \sigma_x^j}{2N} \rangle_{\rho} + \sum_{i<j} \langle \tfrac{\sigma_y^i \sigma_y^j}{2N} \rangle_{\rho} - \tfrac{N-1}{4}.     \label{eq:dicke-fidelity-pauli}
\end{align}
If we repeatedly reprepare $\rho$ and sample $N_s$ values each to estimate $\mathit{MSP}, \langle \tfrac{\sigma_z^i \sigma_z^j}{2N}\rangle, \langle\tfrac{\sigma_x^i \sigma_x^j}{2N}\rangle, \langle\tfrac{\sigma_y^i \sigma_y^j}{2N}\rangle $ as in Equation~\eqref{eq:dicke-fidelity-pauli}, these random samples can be seen as $N_s$ independent samples per each of $1+3\tbinom{N}{2}$ independent random variables $X_i \in [a_i,b_i]$ with $[a_i,b_i] = [0,1]$ for $\mathit{MSP}$ and $[a_i,b_i]=[-\tfrac{1}{2N},\tfrac{1}{2N}]$ for the Pauli terms. 
If, on the other hand, we group commuting measurements in the $Z$-, the $X$- and the $Y$-basis as in Equation~\eqref{eq:dicke-fidelity-commuting}, we get only three independent random variables $X_i$ with $[a_i,b_i] = [0,1+\tfrac{N}{4}]$ for the $Z$-basis term and $[a_i,b_i]=[0,\tfrac{N}{4}]$ both for the $X$- and $Y$-basis terms.

Adding the averages of the $N_s$ sample values per random variable, we get an estimator $\overline{\mathcal{S}}$ for $\langle \mathcal{S} \rangle_{\rho}$. To determine a sufficient number of sample value $N_s$ per random variable, we use Hoeffding's inequality~\cite{hoeffding1963probability} for a $\gamma$-confidence interval of size $\pm\epsilon$ (where the tightness of the bound depends on the actual variance of the random variables):
\begin{align}
    \label{eq:hoeffding}
    \mathrm{Pr}\left[ \lvert N_s\cdot \overline{\mathcal{S}} - N_s \langle \mathcal{S} \rangle_{\rho} \rvert \geq N_s\varepsilon\right] \leq 2 \exp \left( - \frac{2\varepsilon^2 N_s^2}{N_s\sum_i (b_i-a_i)^2} \right) \stackrel{!}{\leq} 1-\gamma
    \quad \Rightarrow\quad
    N_s \geq \frac{\ln(2/(1-\gamma))}{2\varepsilon^2} \cdot \sum_i (b_i-a_i)^2
\end{align}
Plugging in the values from the previous paragraph, we get the following sample complexities:
Either we prepare our state $\rho$ for sampling for each of the $1+3\tbinom{N}{2}\approx 1.5N^2$ terms in Equation~\eqref{eq:dicke-fidelity-pauli} approximately $N_s\approx 92(2.5-\tfrac{3}{2N})$ times,
or we prepare our state $\rho$ for sampling for each of the 3 terms in Equation~\eqref{eq:dicke-fidelity-commuting} approximately $N_s\approx17N^2+46N+92$ times.

\emph{Empirical.}
Although these numbers significantly improve existing sampling overhead estimation techniques for this fidelity lower bound~\cite{somma2006lower}, it is still prohibitive for our larger Dicke state hardware experiments. 
Hence, we ran the Dicke state circuits on the Quantinuum H1-2E simulator first and estimated that $\max(150,4\tbinom{N}{K})$ shots per each of the three measurement basis were sufficient for a $68$\%-confidence interval of size $\pm0.1$. Note that this is smaller than the derived polynomial values for $N_s$ when restricted to $K<N\leq 10$.

\begin{figure*}[t!]
\centering 
\includegraphics[width=0.3\textwidth]{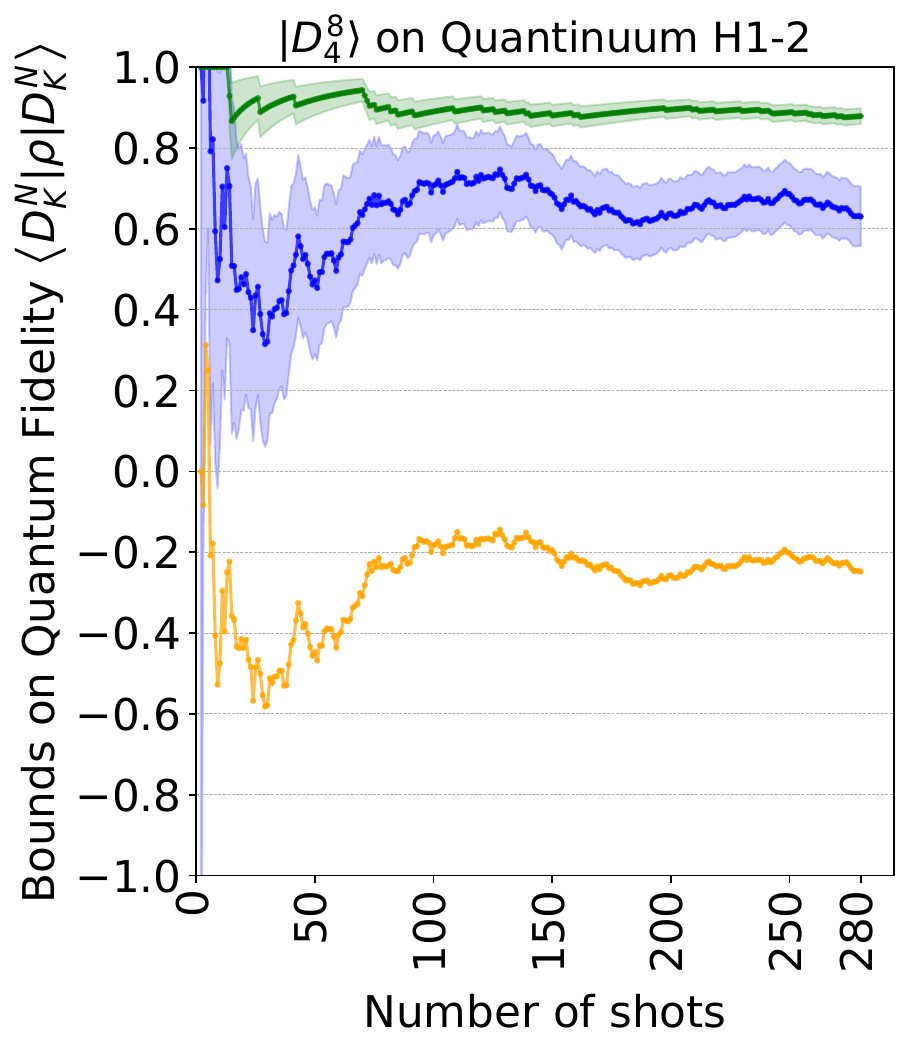}%
\includegraphics[width=0.595\textwidth]{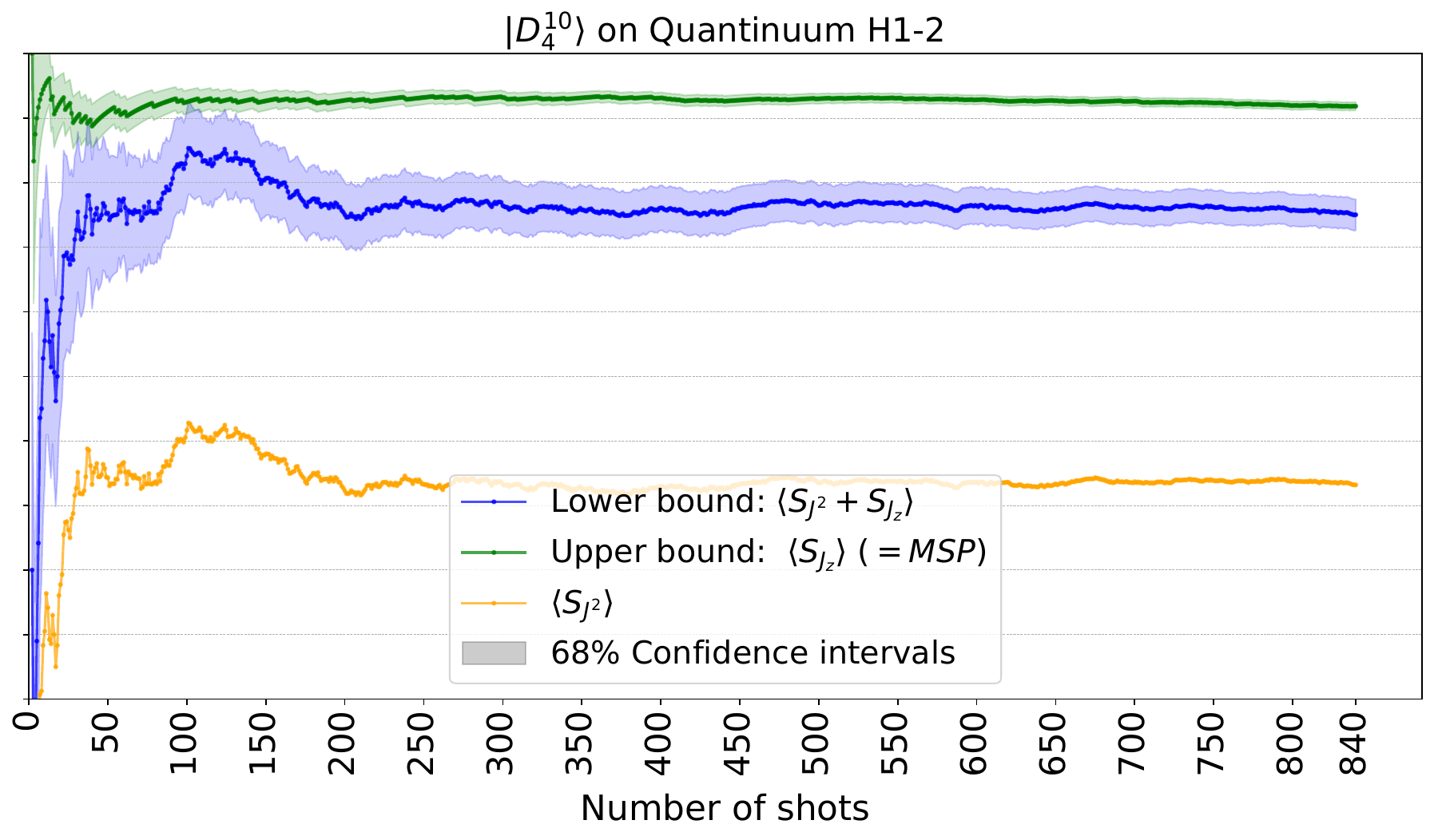}%
\caption{Evolving bounds and confidence intervals on Dicke state fidelity $\dickefidelity{N}{K}$ for 
$\dicke{8}{4}$ and $\dicke{10}{4}$ on Quantinuum H1-2 with increasing shot count. 
Experiments were run for $280$ and $840$ shots for $\dicke{8}{4}$ and $\dicke{10}{4}$, respectively, 
to achieve comparable confidence intervals. 
Plots are scaled (1:3) accordingly.  
}
\label{fig:shots-dicke}
\end{figure*}

Figure~\ref{fig:shots-dicke} shows cumulative measurements of bounds on quantum fidelity $\dickefidelity{N}K{}$. The left plot shows bounds on quantum fidelity for Dicke state $\dicke{8}{4}$ \& the right plot shows the bounds for Dicke state $\dicke{10}{4}$ on H1-2 quantum device. In each plot, the $X$-axis shows the number of shots, and the $Y$-axis shows the cumulative bound estimations. Each plot shows three cumulative lines: lower bound ${\langle \mathcal{S}_{J_z} + \mathcal{S}_{J^2} \rangle}$, upper bound measured success probability and $S_{j^2}$. The plots also show two-sided 68\% confidence intervals below mean and above mean for lower bound and measured success probability $S_{J_z}$. The plots are scaled according to the maximum number of shots used to execute the circuit. In each plot, we find that initially, there is a lot of fluctuation in the lower and upper bound estimations, but the estimations start to get balanced with the increasing number of shots. Similarly, the confidence interval region follows the lower bound estimation and shows how the estimation starts to be stable at the end. The cumulative measurement success probability of the experiment gives an upper bound on quantum fidelity, stabilizing faster than the lower bound.

\section{Fidelity Lower Bounds for GHZ States}
\label{sec:ghz}
We provide experimental lower bound estimations on quantum fidelity for GHZ states $\ghz{N}$ (up to $N=20$). First, we discuss the circuit preparation strategies. Then, we provide lower bound expressions for estimating quantum fidelity in state preparation. Finally, we present experimental results for fidelity estimation in Quantinuum's H1-1 quantum processor.

\subsection{State Preparation}
\label{ghz_prep}
A N-qubit GHZ state is comprised of equal weight superposition of all zeros and all ones defined as,  
	\begin{align}
	    \ghz{N} = \frac{\ket{0}^{\bigotimes N}+\ket{1}^{\bigotimes N}}{\sqrt{2}}.
	\end{align}
The straightforward way to prepare $\ghz{N}$ is first initializing all qubits to $\ket{0}^{\bigotimes N}$. The next step is adding a Hadamard gate $H$ to the first qubit and then consecutively adding a CNOT gate between $i^{th}$ and $(i+1)^{th}$ qubit, where $i = 0$ to $N-1$. This circuit generates a linear time complexity of $N$. Recent work showed that some CNOT gates could be rearranged in GHZ state preparation without affecting the output state, allowing execution of the CNOT gates in parallel~\cite{epfl2019wstate}. Such rearrangement reduces the execution step of the GHZ circuit and generates logarithmic depth GHZ circuits with a time complexity / CNOT depth of $\lceil\log N\rceil$. Figure~\ref{fig:ghz} shows $N = 9$ qubit linear (left) and logarithmic depth (right) GHZ state preparations from initial state $\ket{0}^{\bigotimes N}$. Both circuits contain $N$ gates ($1$ Hadamard and $(N-1)$ CNOTs). The dotted lines in the right plot indicate slices within which the CNOT gates can be performed in parallel. The linear depth circuit (left) requires serial execution of the CNOT gates after the first Hadamard gate that induces linear time complexity of $N = 9$. In the logarithmic depth $\ghz{N}$ circuit, the CNOT gates in each time slice execute in parallel, generating logarithmic complexity of $\lceil\log N\rceil$ = $\lceil\log9\rceil = 4$.

\begin{figure*}[t!]
\centering
\begin{adjustbox}{width=0.99\linewidth}
    		\begin{quantikz}[row sep={24pt,between origins},execute at end picture={}]
	    		\lstick{$q_0\colon\ket{0}$}	& \gate{H}  & \ctrl{8}	& \qw   & \qw\rstick[9]{\rotatebox{90}{$\ghz{9}$}}    &	&	&	&	&	\lstick{$q_0\colon\ket{0}$}	& \gate{H}  & \ctrl{1}  & \qw       & \qw       & \qw       & \qw       & \qw       & \qw       & \qw       & \qw   & \qw\rstick[9]{\rotatebox{90}{$\ghz{9}$}}  &   &   &   &   & \lstick{$q_0\colon\ket{0}$}	& \gate{H}  & \ctrl{1}\slice{(1)}	& \ctrl{(2)}& \qw\slice{(2)}& \ctrl{4}  & \qw       & \qw       & \qw\slice{(3)}& \ctrl{8}\slice{(4)}	& \qw   & \qw\rstick[9]{\rotatebox{90}{$\ghz{9}$}}\\
        		\lstick{$q_1\colon\ket{0}$}	& \qw       & \targ{}	& \qw   & \qw                                         &	&	&	&	&	\lstick{$q_1\colon\ket{0}$}	& \qw       & \targ{0}  & \ctrl{1}  & \qw       & \qw       & \qw       & \qw       & \qw       & \qw       & \qw   & \qw                                         &   &   &   &   & \lstick{$q_1\colon\ket{0}$}	& \qw       & \targ{0}			& \qw       & \ctrl{2}      & \qw       & \ctrl{4}  & \qw       & \qw           & \qw       		& \qw   & \qw           \\
			\lstick{$q_2\colon\ket{0}$}	& \qw       & \targ{}	& \qw   & \qw                                         &	&	&	&	&	\lstick{$q_2\colon\ket{0}$}	& \qw       & \qw       & \targ{0}  & \ctrl{1}  & \qw       & \qw       & \qw       & \qw       & \qw       & \qw   & \qw                                         &   &   &   &   & \lstick{$q_2\colon\ket{0}$}	& \qw       & \qw     			& \targ{0}  & \qw           & \qw       & \qw       & \ctrl{4}  & \qw           & \qw       		& \qw   & \qw           \\
			\lstick{$q_3\colon\ket{0}$}	& \qw       & \targ{}	& \qw   & \qw                                         &	&	&	&	&	\lstick{$q_3\colon\ket{0}$}	& \qw       & \qw       & \qw       & \targ{0}  & \ctrl{1}  & \qw       & \qw       & \qw       & \qw       & \qw   & \qw                                         &   &   &   &   & \lstick{$q_3\colon\ket{0}$}	& \qw       & \qw     			& \qw       & \targ{0}      & \qw       & \qw       & \qw       & \ctrl{4}      & \qw       		& \qw   & \qw           \\
        		\lstick{$q_4\colon\ket{0}$}	& \qw       & \targ{}	& \qw   & \qw                                         &	&	&	&	&	\lstick{$q_4\colon\ket{0}$}	& \qw       & \qw       & \qw       & \qw       & \targ{0}  & \ctrl{1}  & \qw       & \qw       & \qw       & \qw   & \qw                                         &   &   &   &   & \lstick{$q_4\colon\ket{0}$}	& \qw       & \qw     			& \qw       & \qw           & \targ{0}  & \qw       & \qw       & \qw           & \qw       		& \qw   & \qw           \\
        		\lstick{$q_5\colon\ket{0}$}	& \qw       & \targ{}	& \qw   & \qw                                         &	&	&	&	&	\lstick{$q_5\colon\ket{0}$}	& \qw       & \qw       & \qw       & \qw       & \qw       & \targ{0}  & \ctrl{1}  & \qw       & \qw       & \qw   & \qw                                         &   &   &   &   & \lstick{$q_5\colon\ket{0}$}	& \qw       & \qw     			& \qw       & \qw           & \qw       & \targ{0}  & \qw       & \qw           & \qw       		& \qw   & \qw           \\
        		\lstick{$q_6\colon\ket{0}$}	& \qw       & \targ{}	& \qw   & \qw                                         &	&	&	&	&	\lstick{$q_6\colon\ket{0}$}	& \qw       & \qw       & \qw       & \qw       & \qw       & \qw       & \targ{0}  & \ctrl{1}  & \qw       & \qw   & \qw                                         &   &   &   &   & \lstick{$q_6\colon\ket{0}$}	& \qw       & \qw     			& \qw       & \qw           & \qw       & \qw       & \targ{0}  & \qw           & \qw       		& \qw   & \qw           \\        
        		\lstick{$q_7\colon\ket{0}$}	& \qw       & \targ{}	& \qw   & \qw                                         &	&	&	&	&	\lstick{$q_7\colon\ket{0}$}	& \qw       & \qw       & \qw       & \qw       & \qw       & \qw       & \qw       & \targ{0}  & \ctrl{1}  & \qw   & \qw                                         &   &   &   &   & \lstick{$q_7\colon\ket{0}$}	& \qw       & \qw     			& \qw       & \qw           & \qw       & \qw       & \qw       & \targ{0}      & \qw       		& \qw   & \qw           \\        
        		\lstick{$q_8\colon\ket{0}$}	& \qw       & \targ{}	& \qw   & \qw                                         &	&	&	&	&	\lstick{$q_8\colon\ket{0}$}	& \qw       & \qw       & \qw       & \qw       & \qw       & \qw       & \qw       & \qw       & \targ{0}  & \qw   & \qw                                         &   &   &   &   & \lstick{$q_8\colon\ket{0}$}	& \qw       & \qw     			& \qw       & \qw           & \qw       & \qw       & \qw       & \qw           & \targ{0}  		& \qw   & \qw
    		\end{quantikz}
    	\end{adjustbox}
\caption{%
Preparation of a ($N=9$)-qubit GHZ state $\ghz{9}$: (left) Schematics, (center) Linear-depth circuit on LNN connectivity, (right) Logarithmic-depth circuit on full connectivity.
For the linear depth GHZ circuit, all CNOT gates are executed consecutively and generate a linear time $O(N)$ complexity. 
For the logarithmic depth GHZ circuit, the gates leading to (1),(2),(3) \& (4), respectively, can be executed in parallel and generate a logarithmic $O(\log N)$ complexity.%
}
\label{fig:ghz}
\end{figure*}

\subsection{Lower bounds for fidelity estimation}
\label{lb_ghz}
GHZ states belong to the family of stabilizer states used in quantum error correction procedures. From~\cite{somma2006lower}, we find symmetry operators can define that stabilizer states,
    \begin{align}
        \label{ghz-op}
        \hat{\mathcal{O}_s} = +1 \ket{\psi}; s \in [1,S].
    \end{align}
Here, the stabilizer operator $\hat{\mathcal{O}_s}$ are products of Pauli operators that take +1 or -1 as possible eigenvalues. An stabilizer state $\ket{\psi}$ can be uniquely defined by \ref{ghz-op} where $G_s =\{\hat{\mathcal{O}_1},..,\hat{\mathcal{O}_s}\}$ forms the stabilizer group. $G_s$ can also be defined using its $L$ linear independent generators \cite{gottesman1997stabilizer}, $ G_s \equiv (\hat{g_1},...,\hat{g_L})$ where
    \begin{align}
        \hat{g_i}\ket{\psi} = +1\ket{\psi}, i \epsilon [1,L].
    \end{align}
Thus, the state $\ket{\psi}$ can be expressed  as $\ket{\psi} = \ket{g_1 = 1,...,g_L = 1}$ while the density operator $\rho_\psi$ can be defined as
\begin{equation}
   \begin{aligned}
   \label{ghz_den}
   	\rho_\psi	& = \ket{g_1\text{=}1,...,g_L\text{=}1}\bra{g_1\text{=}1,...,g_L\text{=}1}
                 	 = \frac{1}{2^L} \prod_{i=1}^{L} (\hat{g_i} + \mathbb{1}).
   \end{aligned}
\end{equation}

\begin{figure*}[t!]
    \centering 
    \includegraphics[width=0.54\textwidth]{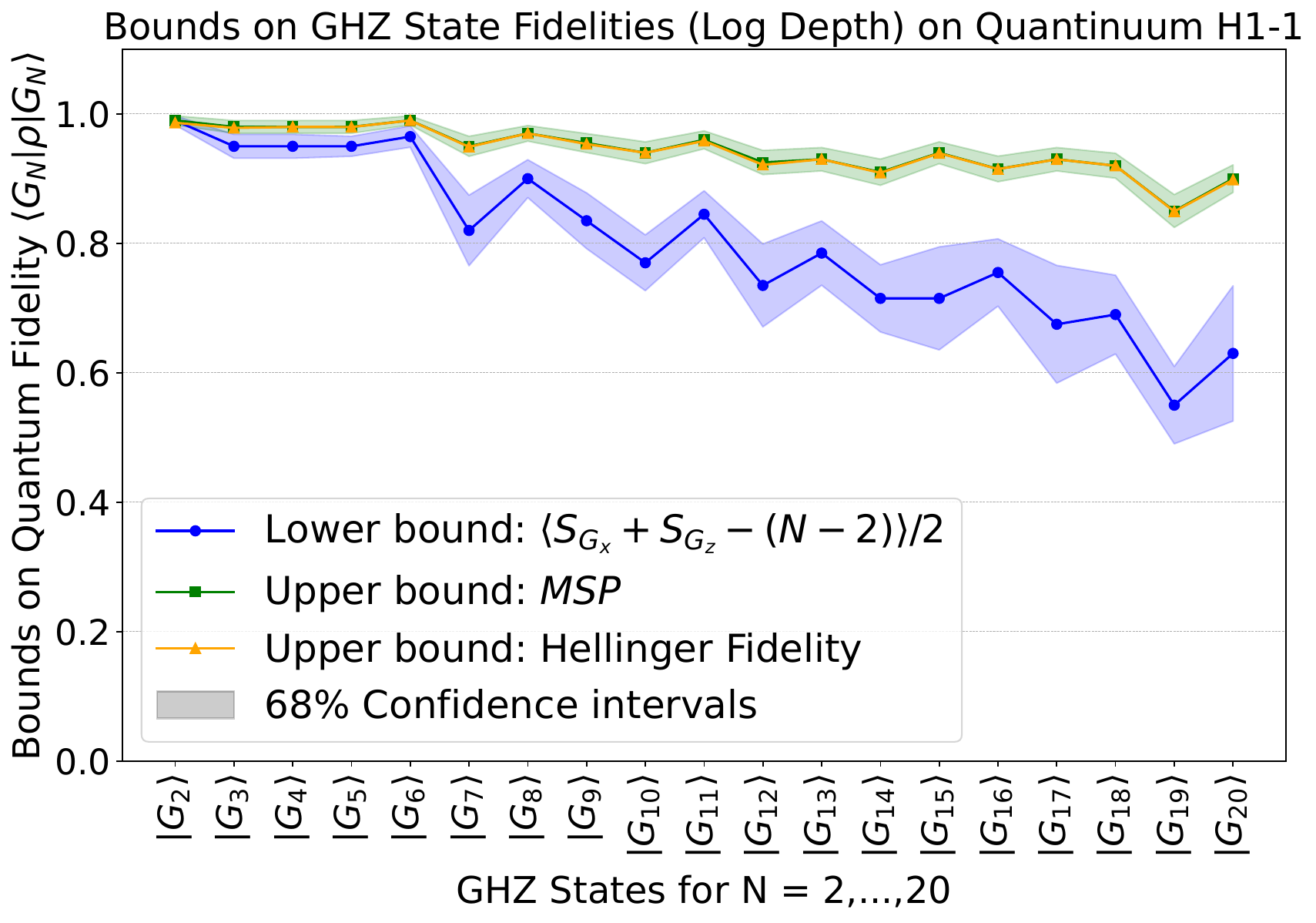}%
    \includegraphics[width=0.50\textwidth]{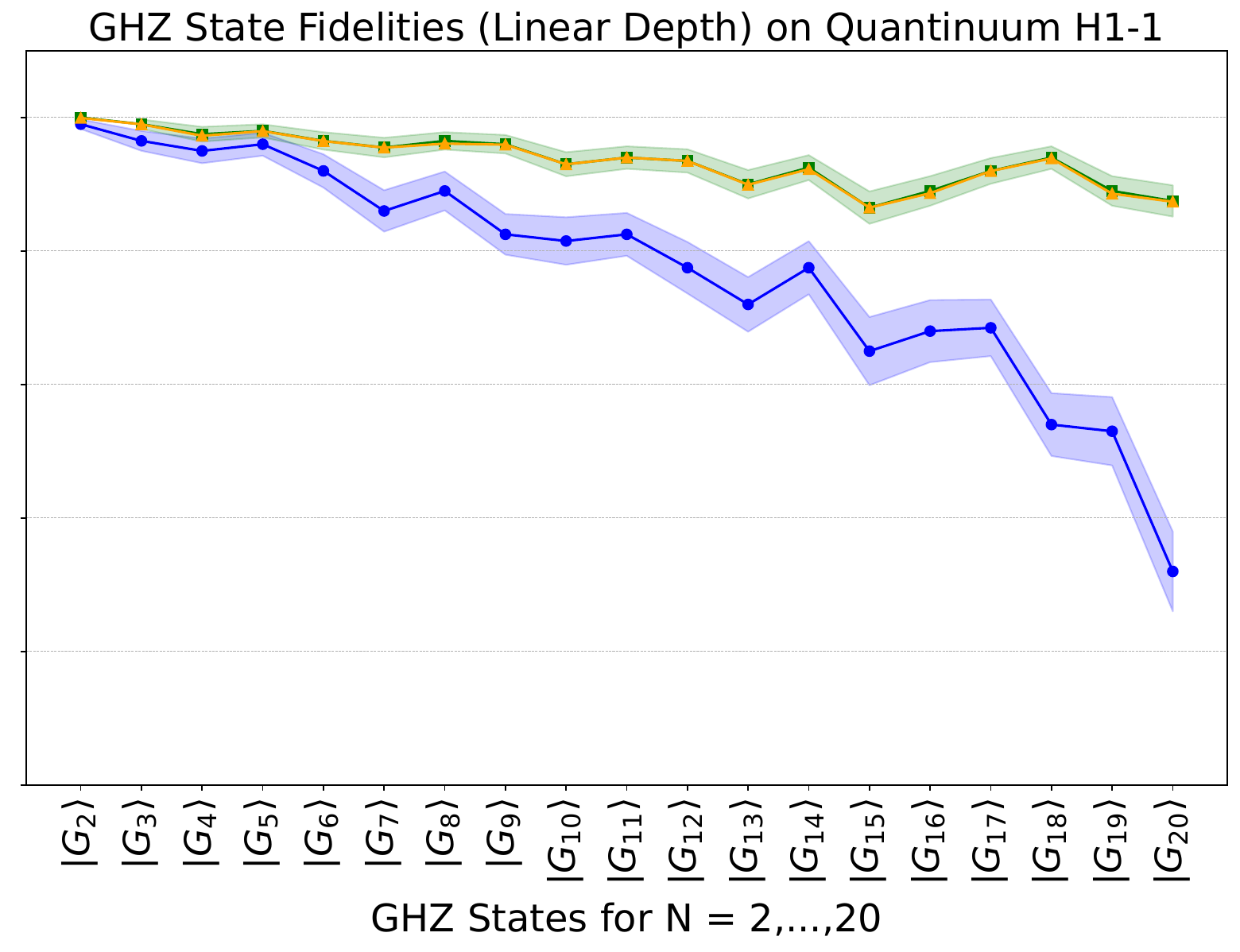}%
    \caption{Upper and lower bounds on quantum fidelity $\ghzfidelity{N}$ for GHZ States $\ghz{N}$ on Quantinuum H1-1 up to $N=20$, including 68\% confidence intervals. 
    The left plot shows results for logarithmic depth state preparation circuits, 
    while the right plot shows results for a linear depth state preparation.
    }
    \label{fig:ghz_all}
\end{figure*}   

The fidelity of a prepared state $\ket{\rho}$ can be measured by computing the expectations listed in \ref{ghz_den}. From~\cite{somma2006lower}, we find that a lower bound on the fidelity can be estimated using the operator $\mathcal{S}_{G_s} = \frac{1}{2}[(\sum_{i=1}^{L} \hat{g_i})-(L-2)]$. When the operator $\mathcal{S}_{G_s}$ is applied to the state $\ket{\psi} = \ket{g_1,...,g_L}$, we get,
   \begin{align}
        \mathcal{S}_{G_s}\ket{g_1,...,g_L} = e_{g_1,...,g_L}\ket{g_1,...,g_L}, (g_i = \pm 1)
   \end{align}
where $e_{1,...,1} = 1$ and $e_{g_1,...g_L \leq 0}$. The lower bound on fidelity between prepared state $\rho$ and expected state $\ghz{N}$ becomes
   \begin{align}
        \label{eq:ghz-fidelity}
       \mathcal{F}(\ghz{N},\rho) \geq{\langle \mathcal{S}_{G_s} \rangle_{\rho}}
   \end{align}
   Using the formulas above for an N-qubit GHZ state $\ghz{N}$, the stabilizer group can be defined as 
   $G_s \equiv (\sigma_x^1\sigma_x^2..\sigma_x^N, \sigma_z^1\sigma_z^2,$\linebreak$\sigma_z^2\sigma_z^3,...,\sigma_z^{N-1}\sigma_z^N)$. We can estimate the lower bounds on GHZ state fidelities using only the Pauli $X$ and $Z$ measurements, and $\mathcal{S}_{G_s}$ will be defined as 
   \begin{align}
        \label{eq:ghz_sg}
        \mathcal{S}_{G_s} & = \frac{1}{2} [\mathcal{S}_{G_x} + \mathcal{S}_{G_z} - (N-2)],
   \end{align}
where $\mathcal{S}_{G_x}$ represents product of Pauli $X$ operator on $i=1$ to $N$ qubits and $\mathcal{S}_{G_z}$ represents sum of the pairwise products of Pauli $Z$ operator on $i^{th}$ and $(i+1)^{th}$ qubit upto $N^{th}$ qubit. For a 4-qubit GHZ state, we get $N=4$ and $L=4$,
   \begin{align}
        \label{ghz_fidelity}
       \mathcal{S}_{G_s} & = \frac{1}{2}[\sigma_x^1\sigma_x^2\sigma_x^3\sigma_x^4+\sigma_z^1\sigma_z^2+\sigma_z^2\sigma_z^3+\sigma_z^3\sigma_z^4-2].
   \end{align}
Basically, we need to compute the expectation of the product of Pauli $X$ measurement on each of the four qubits by taking $+1$ for even parity and $-1$ for odd parity. Similarly, we compute the consecutive pairwise product of Pauli $Z$ measurement on qubits 1 \& 2, 2 \& 3, and 3 \& 4 by checking the pairwise parity. Then, we estimate the fidelity lower bound for the GHZ state by taking the average of those products and subtracting $N-2$ from the sum.

\subsection{Experimental Results}
\label{sec:ghz-exp}
We prepare linear depth and logarithmic depth GHZ states $\ghz{N}$ circuits mentioned in Section~\ref{ghz_prep}. This experiment generates untranspiled QASM circuits for GHZ states $\ghz{N}$, where $2 \leq N \leq 20$. We compute the bounds on quantum fidelity using only two measurements ($X$ and $Z$ basis measurement) instead of measuring $3^N$ state tomography circuits to measure the closeness of prepared GHZ state $\rho$ to the expected state $\ghz{N}$. To estimate lower bound $\langle \mathcal{S}_{G_s} \rangle_{\rho}$, we followed the idea described in Section~\ref{lb_ghz}. We also compute the upper bound measured success probability and the Hellinger fidelity of the prepared states.
    
\subsubsection{Overall Results} Figure~\ref{fig:ghz_all} shows the bounds for GHZ state $\ghz{N}$. The GHZ states are sorted left to right according to CNOT gate counts in untranspiled GHZ circuits. The left plot shows bounds for logarithmic depth GHZ state circuits, while the right plot shows bounds for the linear depth GHZ circuits. Both plots show a lower bound on quantum fidelity and measurement success probability along with a 68\% confidence interval below and above the mean of the distribution and Hellinger fidelity. For both log and linear depth circuits, the lower bound on fidelity decreases almost linearly, albeit with few variations with increasing CNOT counts for higher $N$. The gap between the lower bound and measured success probability increases with increasing CNOT count. The linear circuit shows a faster drop for higher GHZ states than the log depth circuit, which is expected as logarithmic depth circuits require fewer time steps for circuit execution. 
Additionally, the better fidelity lower bounds observed in log-depth GHZ preparation circuits indicate low crosstalk in the system. In both plots, we observe a few irregular ups and downs in the lower bound estimation, which could be because of the statistical noise. The gap between the lower bound and confidence interval region stays almost close in all GHZ states. Our investigation finds that expectation $\mathcal{S}_{G_s}$ (defined as the product of Pauli $X$ operations on each qubit) starts to drop from $\ghz{11}$ and becomes more significant from $\ghz{17}$. There are two flat decreases from linear depth GHZ state $\ghz{17}$ to $\ghz{18}$ and $\ghz{19}$ to $\ghz{20}$,  which the qubit rearrangement in Quantinuum H1-1 device could cause. Furthermore, for both linear and logarithmic depth GHZ circuits, the upper bounds (measured success probability and Hellinger fidelity) overlap. This phenomenon is due to the binary nature of GHZ states and the high fidelity achieved in our state preparations.

\subsubsection{Number of Shots}
\emph{Theoretical.}
As in Section~\ref{sec:dicke-exp}, we first derive a theoretical sampling overhead $N_s$ to get $\gamma=68$\%-confidence interval of size $\pm\varepsilon=0.1$ around our lower bound estimator. We write the fidelity lower bound from Equation~\eqref{eq:ghz-fidelity} as:
\begin{align}
    \mathcal{F}(\ghz{N},\rho) \geq \langle \mathcal{S}_{G_s} \rangle_{\rho} 
    &= \langle \tfrac{\mathcal{S}_{G_x}}{2}\rangle_{\rho} \quad +\ \langle \tfrac{\mathcal{S}_{G_z}}{2} \rangle_{\rho} \qquad\ - \tfrac{N-2}{2}   \label{eq:ghz-fidelity-commuting}   \\
    &= \langle \tfrac{\sigma_x^1\ldots \sigma_x^N}{2}\rangle_{\rho} + \sum_{i=1}^{N-1} \langle \tfrac{\sigma_z^i\sigma_z^{i+1}}{2}\rangle _{\rho}- \tfrac{N-2}{2}.     \label{eq:ghz-fidelity-pauli}
\end{align}
Each of the $N$ Pauli terms in Equation~\eqref{eq:ghz-fidelity-pauli} gives rise to a random variable $X_i \in [a_i,b_i] = [-\tfrac{1}{2},\tfrac{1}{2}]$, while combining commuting measurements in the terms of Equation~\eqref{eq:ghz-fidelity-commuting} results in two random variables $X_i \in [a_i,b_i]$ with $[a_i,b_i] = [-\tfrac{1}{2},\tfrac{1}{2}]$ and $[a_i,b_i] = [-\tfrac{N-1}{2},\tfrac{N-1}{2}]$ for $\langle \tfrac{\mathcal{S}_{G_x}}{2}\rangle$ and $\langle \tfrac{\mathcal{S}_{G_z}}{2} \rangle$, respectively.
Plugging these values into Hoeffding's Inequality~\eqref{eq:hoeffding}, we either have to prepare our state $\rho$ for approximately $N_s\approx92N$ times for each projective measurement of each of the $N$ Pauli terms in Equation~\eqref{eq:ghz-fidelity-pauli}, or approximately $N_s\approx92(N-1)^2+92$ times for both the measurements in the $X$- and the $Z$-bases as in the terms of Equation~\eqref{eq:ghz-fidelity-commuting}.

\emph{Empirical.}
Again, these bounds improve existing results for lower bounds on GHZ fidelity. However, we observed for several GHZ states on the H1-2e emulator that a constant number of $\sim200$ shots in each of the two measurement bases gave good confidence intervals. To see the effect of the number of shots on the quantum fidelity of GHZ states, we execute the GHZ states $\ghz{N}$ where $2 \leq N \leq 20$ for 200 shots on the H1-1 quantum processor. We estimate the cumulative lower bound distribution on quantum fidelity $\ghzfidelity{N}$ from initial to up to $i^{th}$ shot where $i = 2$ to $200$. 

\begin{figure*}[t!]
	      \centering 
        \includegraphics[height=4.6cm]{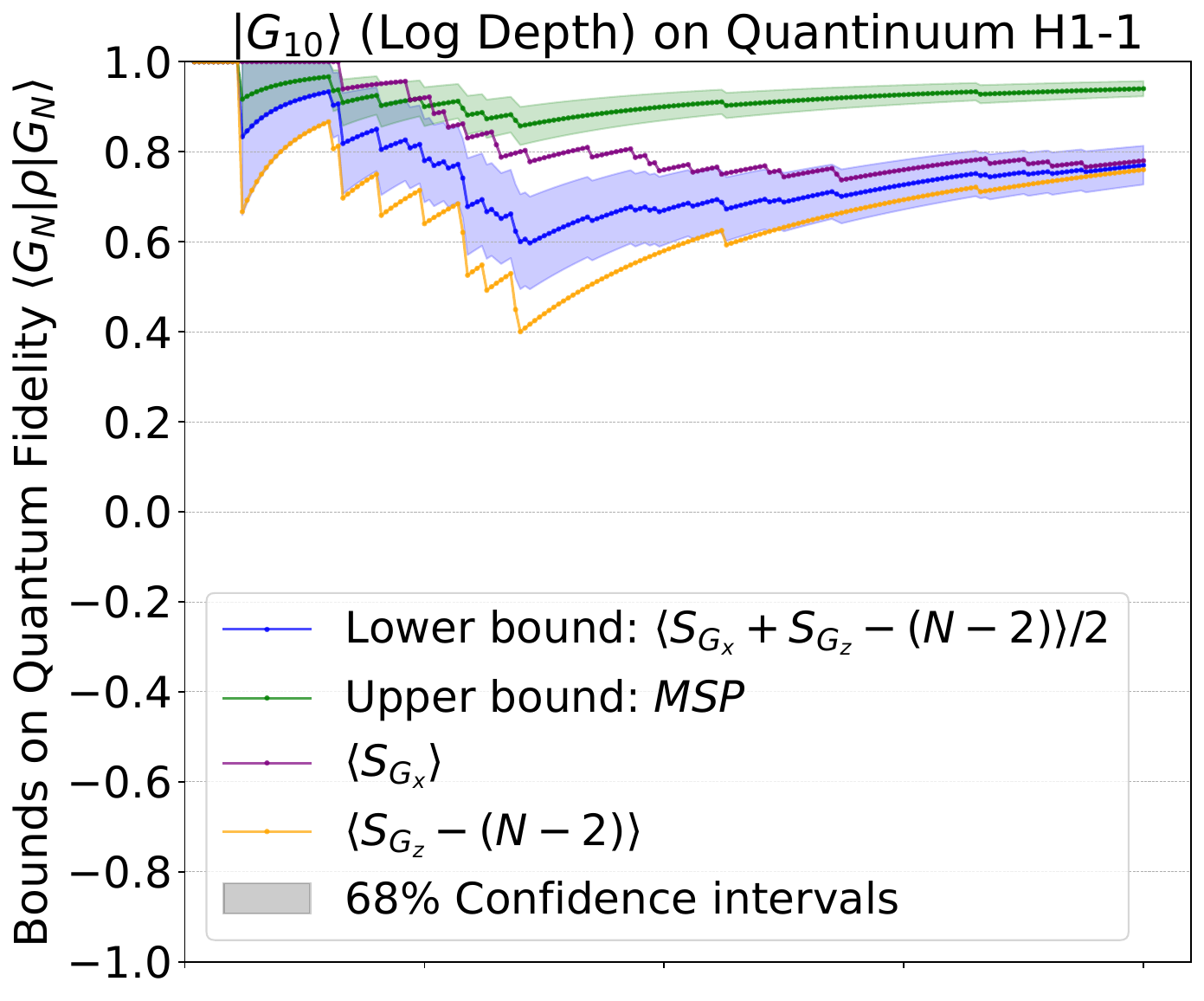}%
        \includegraphics[height=4.6cm]{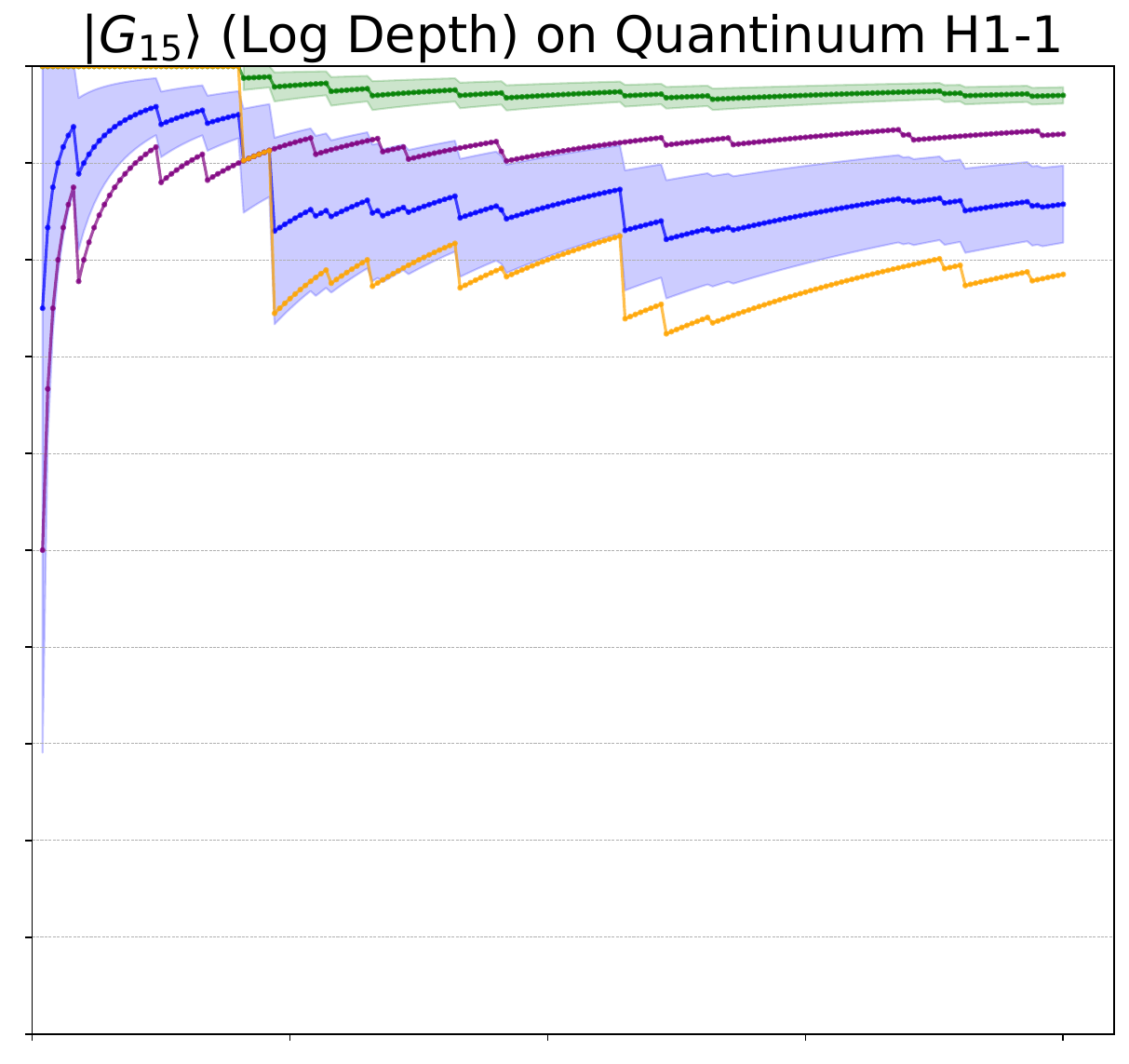}%
        \includegraphics[height=4.6cm]{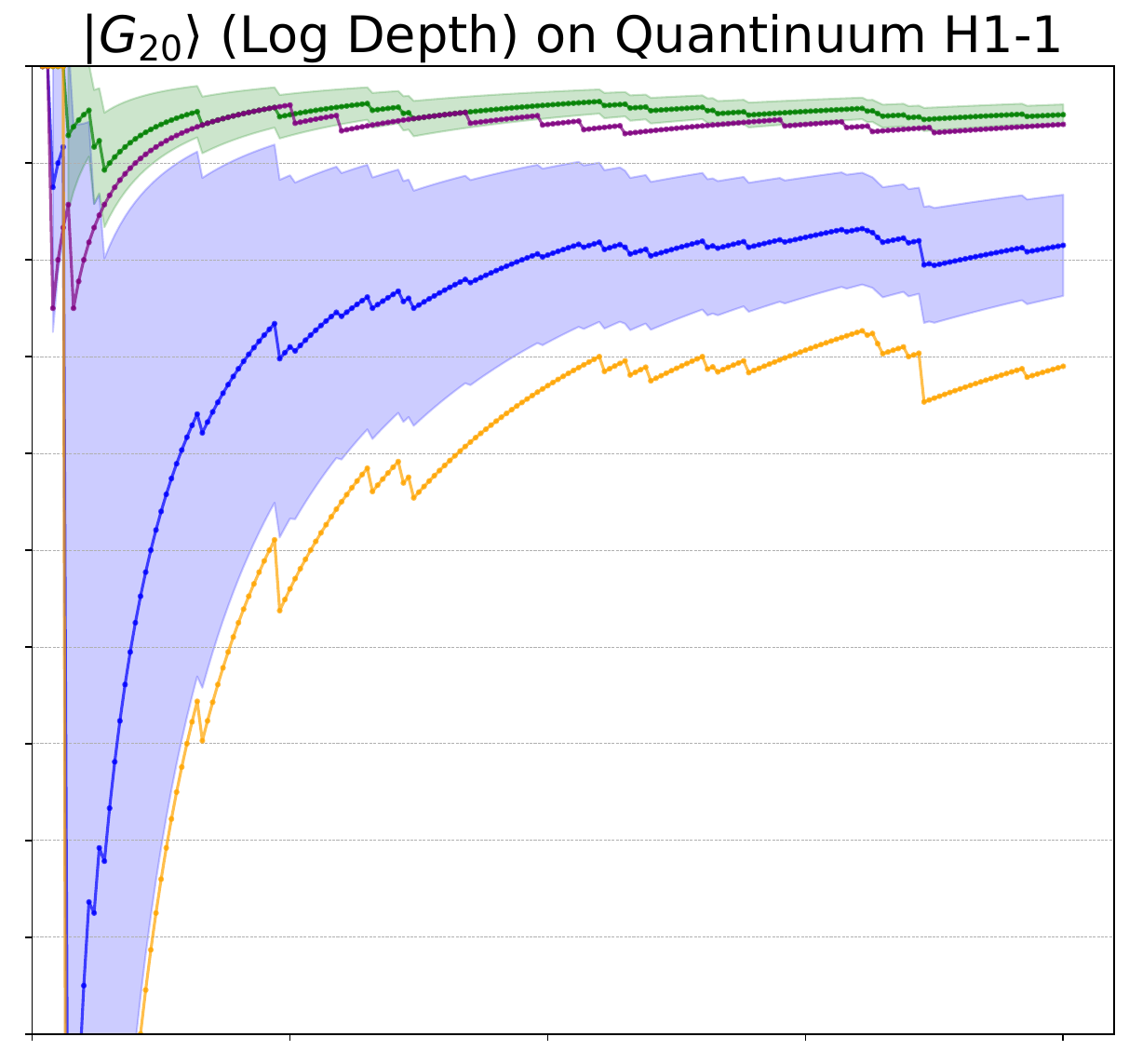}\\
        \includegraphics[height=5.0cm]{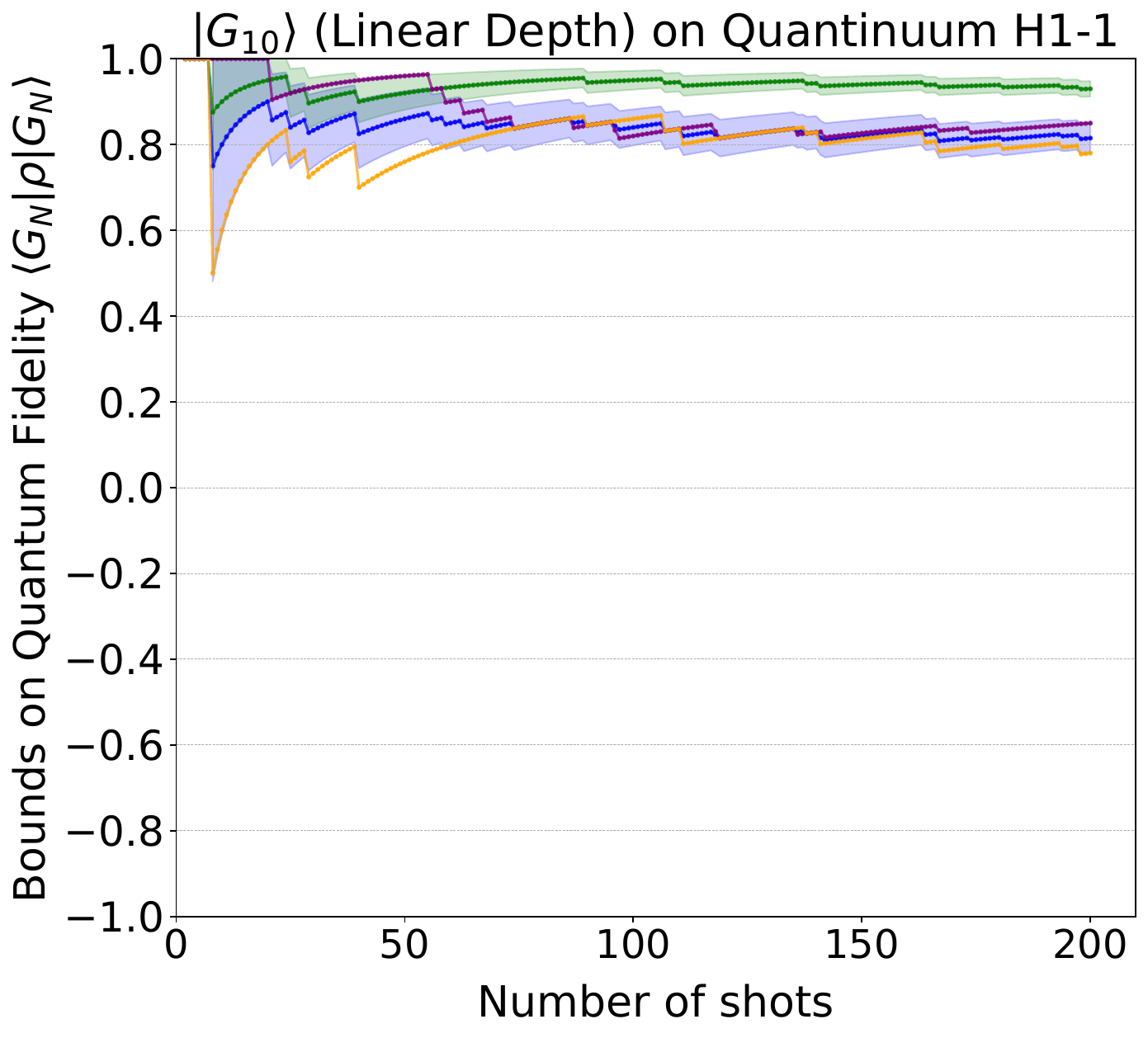}%
        \includegraphics[height=5.0cm]{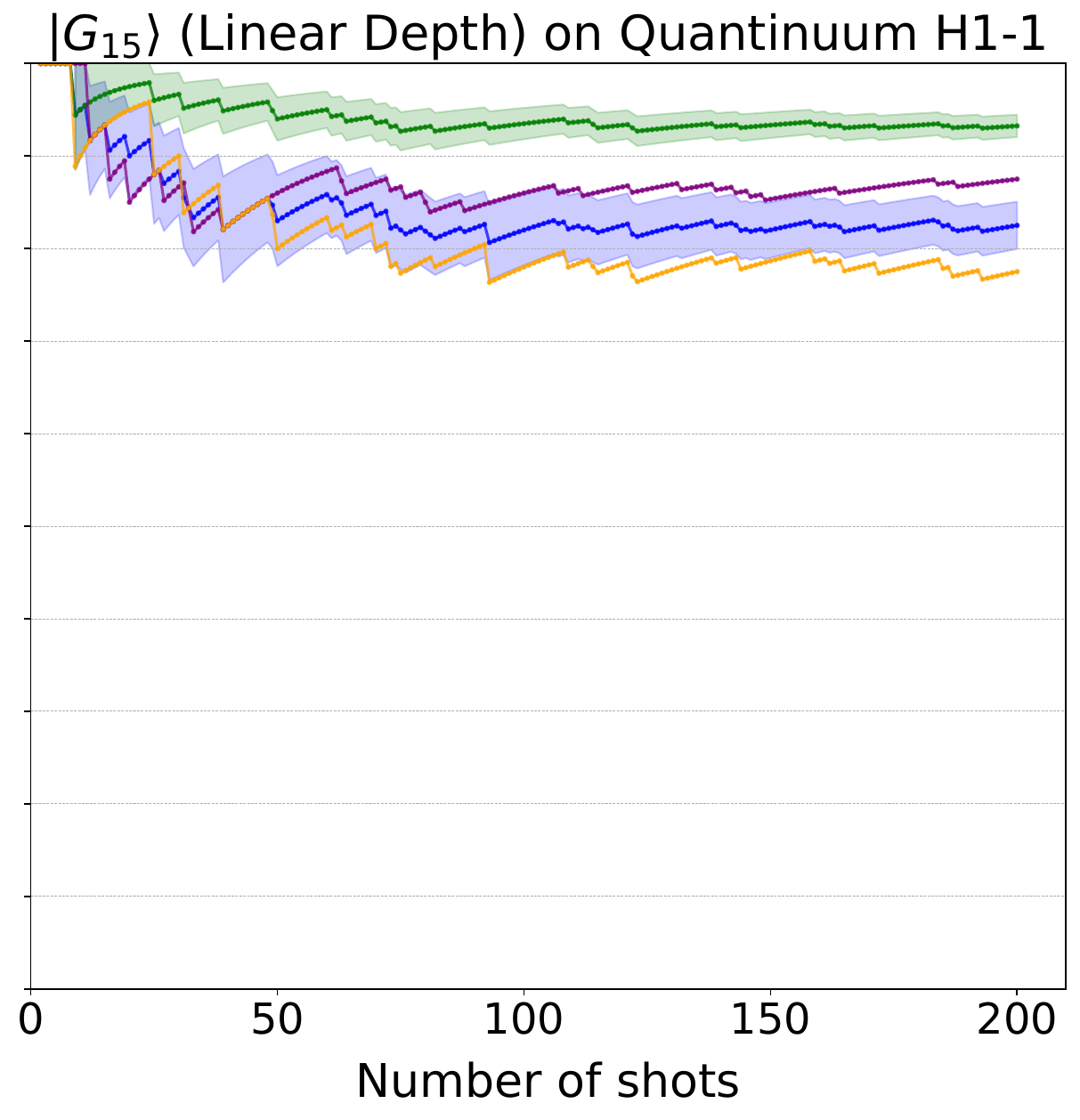}%
        \includegraphics[height=5.0cm]{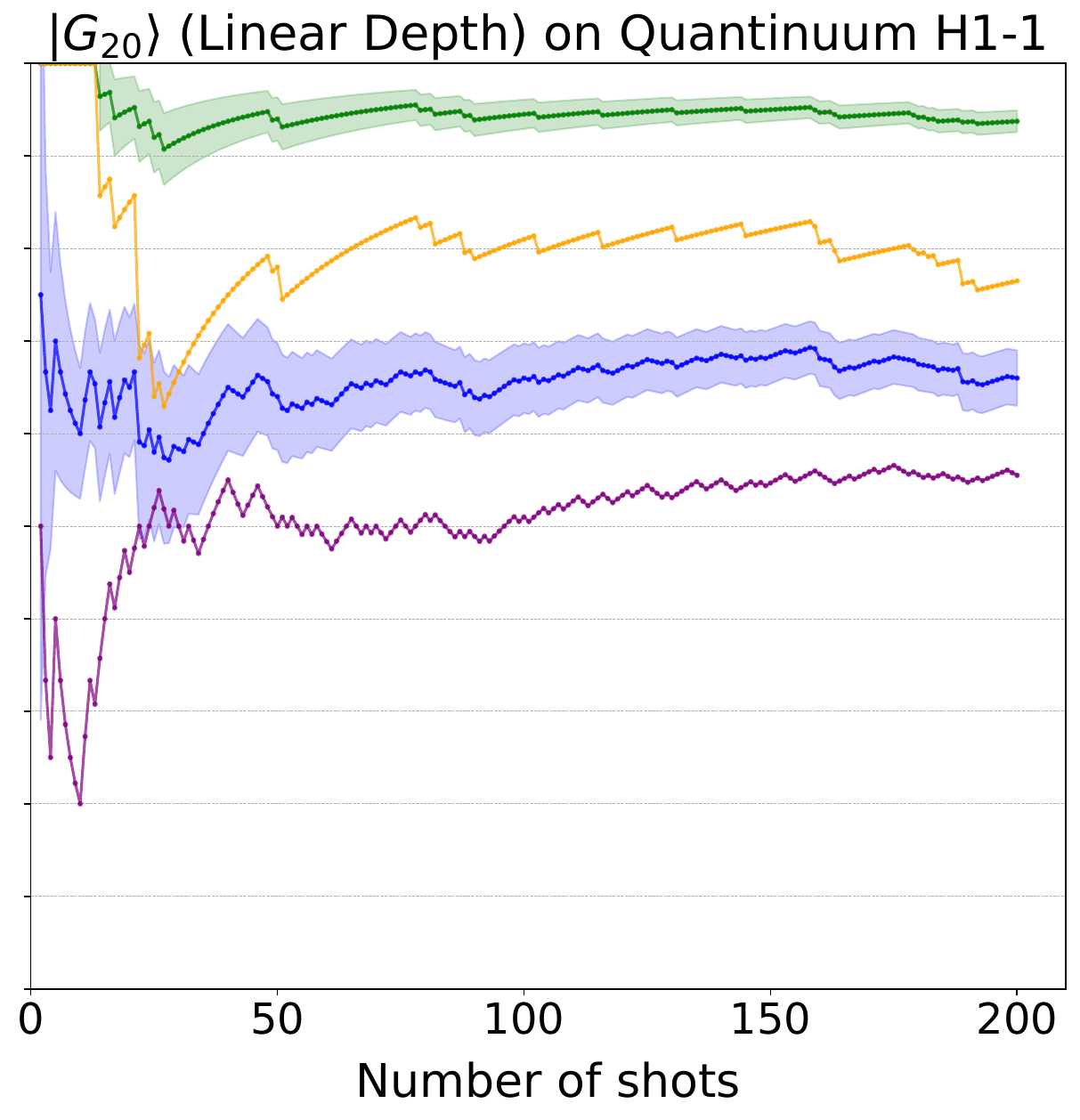}%
        \caption{Evolving bounds on quantum fidelity $\ghzfidelity{N}$ for GHZ States \textit{(left)} $\ghz{10}$, \textit{(center)} $\ghz{15}$ \& \textit{(right)} $\ghz{20}$ on Quantinuum H1-1 over 200 shots.
        \textit{(top row)} Plots show bounds for logarithmic depth GHZ state preparation circuits; \textit{(bottom row)} plots show bounds for linear depth GHZ state preparation. 
        Each plot shows the cumulative progress of lower bound estimation, including confidence intervals.
        All experiments were run back to back in batch mode.
        }
        \label{fig:shots-ghz}
    \end{figure*}

Figure \ref{fig:shots-ghz} shows the estimations of bounds on quantum fidelity $\ghzfidelity{N}$ for GHZ states on Quantinuum H1-1 quantum processor. The upper two plots (left and right) show logarithmic depth $\ghz{10}$, $\ghz{15}$ \& $\ghz{20}$ states, and the bottom two plots  (left and right) show linear depth $\ghz{10}$, $\ghz{15}$ \& $\ghz{20}$ states. Each plot shows lower bound estimation, measured success probability, and expectations of Pauli $X$ \& $Z$ measurements for GHZ state $\ghz{N}$ preparation. The plots also include a 68\% confidence interval region below and above the mean of the lower bound and measured success probability. In each plot, the cumulative measurements start with a lot of fluctuations that become balanced with increasing the number of shots. We observe that all the measurements are more steady in the $\ghz{10}$ plot for both linear and logarithmic depth circuits than $\ghz{15}$ and $\ghz{20}$ plots. We find that the expectation $\mathcal{S}_{G_x}$ (purple line) drops significantly for $\ghz{20}$. It could be possible that larger GHZ states require more shots to become stable. Additionally, we observe that the confidence interval region for the log depth $\ghz{20}$ circuit is less stable than the linear depth circuit. Besides, the upper bound measured success probability stays almost stable in linear and log depth circuits. 

\section{Fidelity Lower Bounds for Approximate Dicke States}
\label{sec:approx-dicke}
Section~\ref{sec:dicke-exp} shows the linear relationship between Dicke state fidelities as the number of qubits ($N$) increases. For higher $N$, direct fidelity computation becomes exponentially expensive in terms of time and resources. Also, the lower bound estimation becomes infeasible because of the statistical noise in the system. As an alternative, preparing an ``approximate'' Dicke state might be another option that yields Dicke states with weaker theoretical fidelities but less noise.

\begin{figure*}[t!]
	\centering
	\begin{adjustbox}{width=0.99\linewidth}
		\begin{quantikz}[row sep={25pt,between origins},execute at end picture={
					\draw[blue,semithick,->]	($(\tikzcdmatrixname-1-10)-(0pt,6pt)$) edge[bend left] ($(\tikzcdmatrixname-1-19)-(30pt,6pt)$);
					\node[fit=(\tikzcdmatrixname-1-22)(\tikzcdmatrixname-9-32),draw,dashed,thick,rounded corners,inner xsep=6pt, inner ysep=10pt, yshift=4pt] {};	
					\node[label={[yshift=-6pt]\large$\ghz{9}$}] at (\tikzcdmatrixname-1-31) {};%
			}]
			\lstick{$q_0\colon\ket{0}$} & \rygate{5/9}{2}{} & \qw\rstick[9]{\rotatebox{90}{Product State: $\ps{9}{4}$}}   	& & & & & & \lstick{$q_0\colon\ket{0}$}	& \qw		& \targ{}	& \targ{}	& \targ{}	& \targ{}	& \targ{}	& \targ{}	& \targ{}	& \targ{}	& \gate{X^{k \pmod 2}}	& \qw	&   &   &\lstick{$\ket{0}$}	& \gate{H}  & \ctrl{1}          & \ctrl{2}  & \qw           & \ctrl{4}  & \qw       & \qw       & \qw           & \ctrl{8}  & \gate{H}   & \gate{X^{k \pmod 2}} 	& \qw\rstick[9]{\rotatebox{90}{$k$ Even/Odd State: $\even{9}$ respectively $\odd{9}$}}\\
			\lstick{$q_1\colon\ket{0}$} & \rygate{5/9}{2}{} & \qw                                                           & & & & & & \lstick{$q_1\colon\ket{0}$}	& \gate{H}	& \ctrl{-1}	& \qw		& \qw		& \qw		& \qw		& \qw		& \qw		& \qw		& \qw			& \qw	&   &   &\lstick{$\ket{0}$}	& \qw       & \targ{0}          & \qw       & \ctrl{2}      & \qw       & \ctrl{4}  & \qw       & \qw           & \qw       & \gate{H}   & \qw       			& \qw                                                   \\
			\lstick{$q_2\colon\ket{0}$} & \rygate{5/9}{2}{} & \qw                                                           & & & & & & \lstick{$q_2\colon\ket{0}$}	& \gate{H}	& \qw		& \ctrl{-2}	& \qw		& \qw		& \qw		& \qw		& \qw		& \qw		& \qw			& \qw	&   &   &\lstick{$\ket{0}$}	& \qw       & \qw               & \targ{0}  & \qw           & \qw       & \qw       & \ctrl{4}  & \qw           & \qw       & \gate{H}   & \qw       			& \qw                                                   \\
			\lstick{$q_3\colon\ket{0}$} & \rygate{5/9}{2}{} & \qw                                                           & & & & & & \lstick{$q_3\colon\ket{0}$}	& \gate{H}	& \qw		& \qw		& \ctrl{-3}	& \qw		& \qw		& \qw		& \qw		& \qw		& \qw			& \qw	&   &   &\lstick{$\ket{0}$}	& \qw       & \qw               & \qw       & \targ{0}      & \qw       & \qw       & \qw       & \ctrl{4}      & \qw       & \gate{H}   & \qw       			& \qw                                                   \\
			\lstick{$q_4\colon\ket{0}$} & \rygate{5/9}{2}{} & \qw                                                           & & & & & & \lstick{$q_4\colon\ket{0}$}	& \gate{H}	& \qw		& \qw		& \qw		& \ctrl{-4}	& \qw		& \qw		& \qw		& \qw		& \qw			& \qw	& = &   &\lstick{$\ket{0}$}	& \qw       & \qw               & \qw       & \qw           & \targ{0}  & \qw       & \qw       & \qw           & \qw       & \gate{H}   & \qw       			& \qw                                                   \\
			\lstick{$q_5\colon\ket{0}$} & \rygate{5/9}{2}{} & \qw                                                           & & & & & & \lstick{$q_5\colon\ket{0}$}	& \gate{H}	& \qw		& \qw		& \qw		& \qw		& \ctrl{-5}	& \qw		& \qw		& \qw		& \qw			& \qw	&   &   &\lstick{$\ket{0}$}	& \qw       & \qw               & \qw       & \qw           & \qw       & \targ{0}  & \qw       & \qw           & \qw       & \gate{H}   & \qw       			& \qw                                                   \\
			\lstick{$q_6\colon\ket{0}$} & \rygate{5/9}{2}{} & \qw                                                           & & & & & & \lstick{$q_6\colon\ket{0}$}	& \gate{H}	& \qw		& \qw		& \qw		& \qw		& \qw		& \ctrl{-6}	& \qw		& \qw		& \qw			& \qw	&   &   &\lstick{$\ket{0}$}	& \qw       & \qw               & \qw       & \qw           & \qw       & \qw       & \targ{0}  & \qw           & \qw       & \gate{H}   & \qw       			& \qw                                                   \\
			\lstick{$q_7\colon\ket{0}$} & \rygate{5/9}{2}{} & \qw                                                           & & & & & & \lstick{$q_7\colon\ket{0}$}	& \gate{H}	& \qw		& \qw		& \qw		& \qw		& \qw		& \qw		& \ctrl{-7}	& \qw		& \qw			& \qw	&   &   &\lstick{$\ket{0}$}	& \qw       & \qw               & \qw       & \qw           & \qw       & \qw       & \qw       & \targ{0}      & \qw       & \gate{H}   & \qw       			& \qw                                                   \\
			\lstick{$q_8\colon\ket{0}$} & \rygate{5/9}{2}{} & \qw                                                           & & & & & & \lstick{$q_8\colon\ket{0}$}	& \gate{H}	& \qw		& \qw		& \qw		& \qw		& \qw		& \qw		& \qw		& \ctrl{-8}	& \qw			& \qw	&   &   &\lstick{$\ket{0}$}	& \qw       & \qw               & \qw       & \qw           & \qw       & \qw       & \qw       & \qw           & \targ{0}  & \gate{H}   & \qw       			& \qw                                                   
		\end{quantikz}
	\end{adjustbox}
	\caption{%
		Preparations of Approximate Dicke States, $N=9$: (left) Product state 
		$\ps{9}{4} = (\surd{\tfrac{5}{9}}\ket{0}+\surd{\tfrac{4}{9}}\ket{1})^{\otimes 9}$, using parallel $R_y(2 \cos^{-1}(\sqrt{5/9}))$ rotations.
		(center) Schematic circuit for Even and Odd Hamming weight state preparation for $K$ even or odd, respectively.
		The circuit adds the Hamming weight parity of a $\ket{+^8}$ state into the first qubit $q_0$. Moving a column of $H$-gates to the right (blue arrow) swaps controls and targets of the CNOT gates.
		This leads to 
		(right) a logarithmic-depth $\ghz{9}$-based state preparation circuit for Even/Odd Hamming weight states $\even{9}$, $\odd{9}$, depending on the parity of $K$. 
	}
	\label{fig:approx-state}
\end{figure*}

\subsection{State Preparation}
We prepare two versions of ``Approximate Dicke states'' with (polynomially) vanishing fidelity to an actual Dicke state $\dicke{N}{K}$:
Product states $\ps{N}{K}$ for small $K$ and Even/Odd Hamming weight states $\even{N}$, $\odd{N}$ for large $K$ with even/odd parity, respectively.
These are defined as:
    \begin{align*}
        \ps{N}{K}   :=  &   \left(\sqrt{1-K/N}\ket{0}+\sqrt{K/N}\ket{1}\right)^{\otimes N}  \\
                    =   &   \sum\nolimits_{i=0}^N \sqrt{\tbinom{N}{i} \left(\tfrac{K}{N}\right)^i \left(\tfrac{N-K}{N}\right)^{N-i}} \dicke{N}{i}
    \end{align*}
    and
    \begin{align*}
        \even{N}    :=  &   \sum\nolimits_{x \in \left\{ 0,1 \right\}^N,\ \hw(x)\equiv 0 \bmod 2}{\ket{x}}  \\
                    =   &   \sum\nolimits_{i=0}^{\lfloor N/2 \rfloor} \sqrt{\tbinom{N}{2i} / 2^{n-1}} \dicke{N}{2i} \\
        \odd{N}     :=  &   \sum\nolimits_{x \in \left\{ 0,1 \right\}^N,\ \hw(x)\equiv 1 \bmod 2}{\ket{x}}  \\
                    =   &   \sum\nolimits_{i=1}^{\lfloor N/2 \rfloor} \sqrt{\tbinom{N}{2i-1} / 2^{n-1}} \dicke{N}{2i-1}  \\
    \end{align*}

Theoretical product state from~\cite{childs2000finding} is a probabilistic approach of preparing $N-qubit$ symmetric product state $(\sqrt{1-K/N}\ket{0}+\sqrt{K/N}\ket{1})^{\bigotimes N}$. The state is prepared by adding rotation $R_y(2\:w
 cos^{-1}(\sqrt{(n-k)/n})$ on each qubit. We then measure them in $X$, $Y$, and $Z$. This yields approximate Dicke state with success probability $\binom{N}{K}(\frac{K}{N})^K(1-\frac{K}{N})^{(N-K)}$. To find approximate Dicke state fidelities, we generate Product state $\ps{N}{K}$ for $N = 2$ to $10$ qubits and $K = 1$ to $N/2$.

Another way of generating approximate Dicke states could be by preparing $N$-qubit odd or even Hamming weight states based on the GHZ circuits discussed in Section~\ref{ghz_prep}. Even Hamming weight N-qubit states are close to Dicke states $\dicke{N}{K}$ with $K$ even and N-qubit odd Hamming weight states give approximate Dicke state $\dicke{N}{K}$ with $K$ odd. Odd/Even Hamming weight states provide Dicke state fidelities of $\binom{N}{K}/2^{N-1}$. $N$-qubit even states are constructed by adding $H^{\bigotimes N}$ at the end of $N$-qubit logarithmic depth GHZ circuits. For $N$-qubit odd states, we just flip one qubit of the even Hamming state. For example, $3$-qubit even Hamming weight state $\sqrt{1/4}(\ket{000}+\ket{011}+\ket{101}+\ket{110})$ are generated from $3$-qubit GHZ states $\sqrt{1/2}(\ket{000}+\ket{111})$. If the first qubit is flipped from $3$-qubit even Hamming weight state, we get odd Hamming weight state $\sqrt{1/4}(\ket{001}+\ket{010}+\ket{100}+\ket{111})$. We get a closer Dicke state with higher fidelity when Hamming weight is higher as Dicke state $\dicke{N}{K}$ has $\binom{N}{K}$ states and odd/even Hamming weight states has $2^{N-1}$ states. When $N$ is higher but $K$ is low, this approximate Dicke state has more non-Dicke states than Dicke states. Therefore, the fidelity gets lower. Thus, we expect higher fidelity with increasing $K/N$. 
We construct odd Hamming weight states $\odd{N}$ for $N = 2$ to $10$ qubits and even Hamming weight states $\even{N}$ for $N = 4$ to $10$ qubits.

\subsection{Lower bounds for Fidelity Estimation}
For N-qubit product state $\ps{N}{K}$, we compute lower bound estimations on Dicke state fidelity using the expressions from Section~\ref{lb-dicke}. Similarly, we estimate Dicke state fidelity for odd and even Hamming weight states. When $K$ is odd, we estimate a lower bound on fidelity for odd Hamming weight state $\odd{N}$, and for even $K$, we compute a lower bound on fidelity for even Hamming weight state $\even{N}$.

\begin{figure*}[t!]
    \centering
    \includegraphics[width=0.99\linewidth]{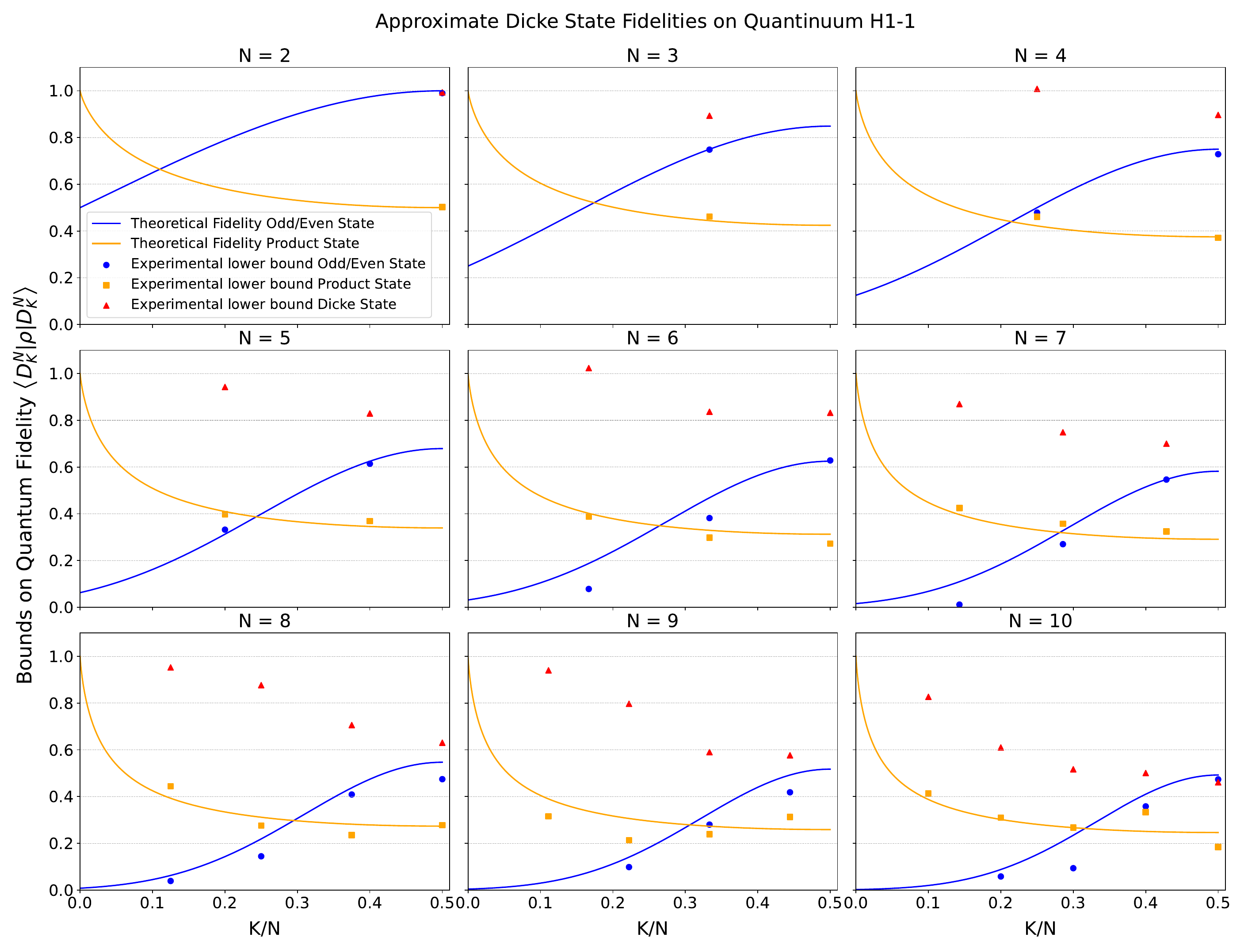}
        \caption{Approximate Dicke State (Product State and Odd/Even State) fidelities with Dicke State and their Lower Bound estimation for $N = 2$ to $10$. Each plot shows theoretical and experimental fidelity for approximate Dicke states along with the lower bound estimations of the prepared Dicke states. Data points are plotted along the $x$-axes according to the ratio $K/N$; theoretical fidelities are plotted as continuous lines to visualize the trend when going to higher $N$. 
    }
    \label{fig:approx_dicke}
\end{figure*}
\subsection{Experimental Results}
We prepare untranspiled circuits for product states $\ps{N}{K}$ where $1 \leq 2K \leq N \leq 10$, odd Hamming weight states $\odd{N}$ where $2\leq N \leq 10$ and even Hamming weight states $\even{N}$ where $4\leq N \leq 10$. Figure~\ref{fig:approx-state} shows the preparation of approximate Dicke States for $N = 9$. We execute the circuits in Quantinuum H1-1 quantum processor to get $X$,$Y$, and $Z$ basis measurements similar to Dicke state circuits in Section~\ref{sec:dicke-exp}. We then compute the experimental lower bound on Dicke state fidelities for product states and odd/even states. Figure~\ref{fig:approx_dicke} shows the experimental bounds on Dicke state fidelities using the approximate Dicke state circuits. The plot shows bounds on quantum fidelity $\dickefidelity{N}{K}$ for $N = 2$ to $N = 10$ qubits. In each plot, the $X$ axis shows values for $K/N$, and the $Y$ axis shows bounds on fidelity. Each plot contains two lines showing theoretical fidelity of odd/even state and product state preparation with Dicke state. In each plot, the bounds (theoretical and experimental) increase for odd/even states with higher $K/N$ but decrease for product states. Experimental lower bounds are also shown for the preparation of odd/even state, product state, and Dicke state (from Section~\ref{sec:dicke-exp}). We find that the experimental lower bounds of the product state closely follow theoretical fidelities in all the plots ($N=2$ to $10$). However, the experimental lower bounds of odd/even states are not close but follow theoretical fidelities. We also observe that the approximate lower bounds of product and odd/even states are getting close to Dicke state lower bounds with higher $N$. For $N = 10$ and $K = 5$, the odd/even state lower bounds match Dicke state lower bound estimations. Thus, we can get a good approximation of lower bounds on Dicke state fidelities from just preparing those approximate states for higher Dicke states.

\section{Conclusion}
\label{sec:discussion}
\label{sec:conclusion}
In this work, we compute lower bounds of quantum fidelity for highly entangled states using only three measurement settings (X, Y, and Z basis) for Dicke State and only two measurement settings (X and Z basis) for GHZ States. Additionally, we show lower bounds of quantum fidelity for approximate Dicke states in terms of product states and odd/even states using only three measurement settings, the same as Dicke states. Our experiments in Sections~\ref{sec:dicke} and~\ref{sec:ghz} show that our computed lower bounds are consistent and follow the upper bound measured success probability. This lower bound computation technique efficiently verifies prepared entangled states of higher qubits since it requires very few measurements. 

The exact fidelity computation using full-state tomography  ($3^N$ measurements) would be very inefficient, expensive, and time-consuming for larger states. To illustrate this, consider that computing the exact fidelity of Dicke states $\dicke{6}{3}$ and $\dicke{10}{5}$ would require around 11.5k HQC and 15M HQC, respectively, to execute on Quantinuum H-system backend. A standard subscription of  Quantinuum H1+H2 through Microsoft Azure is currently (2023) advertised for USD 135k for a monthly HQC allocation of 10k~\cite{azure-quantinuum}. Thus, the exact fidelity computation of Dicke state $\dicke{6}{3}$ would take around one month for USD 135k, and $\dicke{10}{5}$ would take around 125 years at a cost of about USD 200M. Instead, we showed lower bounds of Dicke states $\dicke{6}{3}$ and $\dicke{10}{5}$ with a reasonable confidence interval, and that together required only 814 HQC.

In conclusion, we have demonstrated good fidelity values for the Dicke and GHZ, highly entangled quantum states, prepared on Quantinuum H1 ion trap systems. Since direct fidelity estimation through full-state tomography does not scale, we compute lower bounds on the fidelity with exponentially reduced overhead for those states. For the first time and after 15 years of its theoretical introduction, we show that these bounds are tight enough for meaningful fidelity estimations on the newest NISQ devices. We give state preparation fidelity lower bounds of 0.46 for the Dicke State $\dicke{10}{5}$ and 0.73 for the GHZ State $\ghz{20}$ which match or exceed exact fidelity records recently achieved on superconducting systems for the much smaller states $\dicke{6}{3}$ and $\ghz{5}$, respectively. This improved performance is due to improvements in NISQ hardware and efficient implementations of the latest circuits. Additionally, we show GHZ-based approximate states, \ie, product state and odd/even states can result in a good approximation of large Dicke states $\dicke{N}{N/2}$.
\section{Acknowledgement}
\label{sec:ack}

The US Department of Energy partially supported this work through the Los Alamos National Laboratory, which is operated by Triad National Security, LLC, for the National Nuclear Security Administration of the US Department of Energy (Contract No. 89233218CNA000001). The research presented in this article was supported by the NNSA's Advanced Simulation and Computing Beyond Moore's Law Program at Los Alamos National Laboratory. This work has been assigned LANL technical report number LA-UR-23-31364.
Ms.\ Aktar is supported by a College of Engineering at New Mexico State University Fellowship.
This research used resources from the Oak Ridge Leadership Computing Facility, a DOE Office of Science User Facility supported under Contract DE-AC05-00OR22725. The Oak Ridge Leadership Computing Facility provided access to the Quantinuum H1 computer. 
\bibliographystyle{plainurl} 
\bibliography{lowerbound_arxiv-2025}

\begin{thebibliography}{10}

\bibitem{quantinuum}
Hardware: System model h1, 2023.
\newblock URL: \url{https://www.quantinuum.com/hardware/h1}.

\bibitem{10.1145/3587135.3592197}
Shamminuj Aktar, Abdel-Hameed~A. Badawy, Andreas B\"{a}rtschi, and Stephan
  Eidenbenz.
\newblock Scalable experimental bounds for dicke and ghz states fidelities.
\newblock In {\em Proceedings of the 20th ACM International Conference on
  Computing Frontiers}, CF '23, page 176–184, New York, NY, USA, 2023.
  Association for Computing Machinery.
\newblock \href {https://doi.org/10.1145/3587135.3592197}
  {\path{doi:10.1145/3587135.3592197}}.

\bibitem{aktar2022divide}
Shamminuj Aktar, Andreas B{\"a}rtschi, Abdel-Hameed~A Badawy, and Stephan
  Eidenbenz.
\newblock {A Divide-and-Conquer Approach to Dicke State Preparation}.
\newblock {\em IEEE Transactions on Quantum Engineering}, 3:1--16, 2022.
\newblock \href {http://arxiv.org/abs/2112.12435} {\path{arXiv:2112.12435}},
  \href {https://doi.org/10.1109/TQE.2022.3174547}
  {\path{doi:10.1109/TQE.2022.3174547}}.

\bibitem{altepeter2004quantum}
Joseph~B. Altepeter, Daniel~F.V. James, and Paul~G. Kwiat.
\newblock {\em Quantum State Estimation}, chapter 4 Qubit Quantum State
  Tomography, pages 113--145.
\newblock Springer, 2004.
\newblock \href {https://doi.org/10.1007/978-3-540-44481-7_4}
  {\path{doi:10.1007/978-3-540-44481-7_4}}.

\bibitem{baertschi2019deterministic}
Andreas B{\"a}rtschi and Stephan Eidenbenz.
\newblock {Deterministic Preparation of Dicke States}.
\newblock In {\em 22nd International Symposium on Fundamentals of Computation
  Theory (FCT'19)}, pages 126--139, Cham, 2019. Springer.
\newblock \href {http://arxiv.org/abs/1904.07358} {\path{arXiv:1904.07358}},
  \href {https://doi.org/10.1007/978-3-030-25027-0_9}
  {\path{doi:10.1007/978-3-030-25027-0_9}}.

\bibitem{baertschi2022shortdepth}
Andreas B{\"{a}}rtschi and Stephan Eidenbenz.
\newblock {Short-Depth Circuits for Dicke State Preparation}.
\newblock In {\em IEEE International Conference on Quantum Computing \&
  Engineering (QCE'22)}, pages 87--96, USA, 2022. IEEE.
\newblock \href {http://arxiv.org/abs/2207.09998} {\path{arXiv:2207.09998}},
  \href {https://doi.org/10.1109/QCE53715.2022.00027}
  {\path{doi:10.1109/QCE53715.2022.00027}}.

\bibitem{blumekohout2012robust}
Robin Blume-Kohout.
\newblock Robust error bars for quantum tomography.
\newblock {\em arXiv preprint}, 2012.
\newblock \href {http://arxiv.org/abs/1202.5270} {\path{arXiv:1202.5270}},
  \href {https://doi.org/10.48550/arXiv.1202.5270}
  {\path{doi:10.48550/arXiv.1202.5270}}.

\bibitem{buzek1998reconstruction}
V.~Bu\v{z}ek, R.~Derka, G.~Adam, and P.L. Knight.
\newblock Reconstruction of quantum states of spin systems: From quantum
  bayesian inference to quantum tomography.
\newblock {\em Annals of Physics}, 266(2):454--496, 1998.
\newblock \href {http://arxiv.org/abs/quant-ph/9701029}
  {\path{arXiv:quant-ph/9701029}}, \href
  {https://doi.org/10.1006/aphy.1998.5802} {\path{doi:10.1006/aphy.1998.5802}}.

\bibitem{childs2000finding}
Andrew~M. Childs, Edward Farhi, Jeffrey Goldstone, and Sam Gutmann.
\newblock Finding cliques by quantum adiabatic evolution.
\newblock {\em Quantum Info. Comput.}, 2(3):181–191, apr 2002.
\newblock \href {http://arxiv.org/abs/quant-ph/0012104}
  {\path{arXiv:quant-ph/0012104}}.

\bibitem{christandl2012reliable}
Matthias Christandl and Renato Renner.
\newblock {Reliable Quantum State Tomography}.
\newblock {\em Physical Review Letters}, 109:120403, Sep 2012.
\newblock \href {http://arxiv.org/abs/1108.5329} {\path{arXiv:1108.5329}},
  \href {https://doi.org/10.1103/PhysRevLett.109.120403}
  {\path{doi:10.1103/PhysRevLett.109.120403}}.

\bibitem{cross2017open}
Andrew~W Cross, Lev~S Bishop, John~A Smolin, and Jay~M Gambetta.
\newblock {Open quantum assembly language}.
\newblock {\em arXiv preprint}, 2017.
\newblock \href {http://arxiv.org/abs/1707.03429} {\path{arXiv:1707.03429}}.

\bibitem{epfl2019wstate}
{Cruz, Diogo and Fournier, Romain and Gremion, Fabien and Jeannerot, Alix and
  Komagata, Kenichi and Tosic, Tara and Thiesbrummel, Jarla and Chan, Chun Lam
  and Macris, Nicolas and Dupertuis, Marc-André and Javerzac-Galy, Clément}.
\newblock {Efficient Quantum Algorithms for GHZ and W States, and
  Implementation on the IBM Quantum Computer}.
\newblock {\em Advanced Quantum Technologies}, 2(5-6):1900015, 2019.
\newblock \href {http://arxiv.org/abs/1807.05572} {\path{arXiv:1807.05572}},
  \href {https://doi.org/10.1002/qute.201900015}
  {\path{doi:10.1002/qute.201900015}}.

\bibitem{dariano2003quantumtomography}
G.~Mauro D'Ariano, Matteo G.~A. Paris, and Massimiliano~F. Sacchi.
\newblock {\em Advances in Imaging and Electron Physics}, volume 128, chapter
  {Quantum Tomography}, pages 205--308.
\newblock Elsevier, 2003.
\newblock \href {http://arxiv.org/abs/quant-ph/0302028}
  {\path{arXiv:quant-ph/0302028}}.

\bibitem{elben2022randomized}
Andreas Elben, Steven~T Flammia, Hsin-Yuan Huang, Richard Kueng, John Preskill,
  Beno{\^\i}t Vermersch, and Peter Zoller.
\newblock {The randomized measurement toolbox}.
\newblock {\em Nature Reviews Physics}, 5(1):9--24, dec 2022.
\newblock URL: \url{https://doi.org/10.1038%2Fs42254-022-00535-2}, \href
  {http://arxiv.org/abs/2203.11374} {\path{arXiv:2203.11374}}, \href
  {https://doi.org/10.1038/s42254-022-00535-2}
  {\path{doi:10.1038/s42254-022-00535-2}}.

\bibitem{flammia2011direct}
Steven~T Flammia and Yi-Kai Liu.
\newblock {Direct fidelity estimation from few Pauli measurements}.
\newblock {\em {Physical Review Letters}}, 106(23):230501, 2011.
\newblock \href {http://arxiv.org/abs/1104.4695} {\path{arXiv:1104.4695}},
  \href {https://doi.org/10.1103/PhysRevLett.106.230501}
  {\path{doi:10.1103/PhysRevLett.106.230501}}.

\bibitem{gottesman1997stabilizer}
Daniel Gottesman.
\newblock {\em {Stabilizer codes and quantum error correction. Caltech Ph. D}}.
\newblock PhD thesis, Thesis, eprint, 1997.
\newblock \href {http://arxiv.org/abs/quant-ph/9705052}
  {\path{arXiv:quant-ph/9705052}}.

\bibitem{guhne2007toolbox}
Otfried G{\"u}hne, Chao-Yang Lu, Wei-Bo Gao, and Jian-Wei Pan.
\newblock {Toolbox for entanglement detection and fidelity estimation}.
\newblock {\em Physical Review A}, 76(3):030305, 2007.
\newblock \href {http://arxiv.org/abs/0706.2432} {\path{arXiv:0706.2432}},
  \href {https://doi.org/10.1103/PhysRevA.76.030305}
  {\path{doi:10.1103/PhysRevA.76.030305}}.

\bibitem{hoeffding1963probability}
Wassily Hoeffding.
\newblock {Probability Inequalities for Sums of Bounded Random Variables}.
\newblock {\em Journal of the American Statistical Association},
  58(301):13--30, 1963.
\newblock \href {https://doi.org/10.1080/01621459.1963.10500830}
  {\path{doi:10.1080/01621459.1963.10500830}}.

\bibitem{huang2020predicting}
Hsin-Yuan Huang, Richard Kueng, and John Preskill.
\newblock Predicting many properties of a quantum system from very few
  measurements.
\newblock {\em Nature Physics}, 16(10):1050--1057, 2020.
\newblock \href {http://arxiv.org/abs/2002.08953} {\path{arXiv:2002.08953}},
  \href {https://doi.org/10.1038/s41567-020-0932-7}
  {\path{doi:10.1038/s41567-020-0932-7}}.

\bibitem{james2004measurement}
Daniel F.~V. James, Paul~G. Kwiat, William~J. Munro, and Andrew~G. White.
\newblock Measurement of qubits.
\newblock {\em Physical Review A}, 64:052312, Oct 2001.
\newblock \href {http://arxiv.org/abs/quant-ph/0103121}
  {\path{arXiv:quant-ph/0103121}}, \href
  {https://doi.org/10.1103/PhysRevA.64.052312}
  {\path{doi:10.1103/PhysRevA.64.052312}}.

\bibitem{jiang2020towards}
Xinhe Jiang, Kun Wang, Kaiyi Qian, Zhaozhong Chen, Zhiyu Chen, Liangliang Lu,
  Lijun Xia, Fangmin Song, Shining Zhu, and Xiaosong Ma.
\newblock {Towards the standardization of quantum state verification using
  optimal strategies}.
\newblock {\em npj Quantum Information}, 6(1):1--8, 2020.
\newblock \href {http://arxiv.org/abs/2002.00640} {\path{arXiv:2002.00640}},
  \href {https://doi.org/10.1038/s41534-020-00317-7}
  {\path{doi:10.1038/s41534-020-00317-7}}.

\bibitem{jozsa1994fidelity}
Richard Jozsa.
\newblock {Fidelity for mixed quantum states}.
\newblock {\em Journal of Modern Optics}, 41(12):2315--2323, 1994.
\newblock \href {https://doi.org/10.1080/09500349414552171}
  {\path{doi:10.1080/09500349414552171}}.

\bibitem{li2019efficient}
Zihao Li, Yun-Guang Han, and Huangjun Zhu.
\newblock {Efficient verification of bipartite pure states}.
\newblock {\em Physical Review A}, 100(3):032316, 2019.
\newblock \href {http://arxiv.org/abs/1901.09783} {\path{arXiv:1901.09783}},
  \href {https://doi.org/10.1103/PhysRevA.100.032316}
  {\path{doi:10.1103/PhysRevA.100.032316}}.

\bibitem{li2020optimal}
Zihao Li, Yun-Guang Han, and Huangjun Zhu.
\newblock {Optimal verification of greenberger-horne-zeilinger states}.
\newblock {\em {Physical Review Applied}}, 13(5):054002, 2020.
\newblock \href {http://arxiv.org/abs/1909.08979} {\path{arXiv:1909.08979}},
  \href {https://doi.org/10.1103/PhysRevApplied.13.054002}
  {\path{doi:10.1103/PhysRevApplied.13.054002}}.

\bibitem{liu2019efficient}
Ye-Chao Liu, Xiao-Dong Yu, Jiangwei Shang, Huangjun Zhu, and Xiangdong Zhang.
\newblock {Efficient verification of Dicke states}.
\newblock {\em Physical Review Applied}, 12(4):044020, 2019.
\newblock \href {http://arxiv.org/abs/1904.01979} {\path{arXiv:1904.01979}},
  \href {https://doi.org/10.1103/PhysRevApplied.12.044020}
  {\path{doi:10.1103/PhysRevApplied.12.044020}}.

\bibitem{mukherjee2020preparing}
Chandra~Sekhar Mukherjee, Subhamoy Maitra, Vineet Gaurav, and Dibyendu Roy.
\newblock {Preparing Dicke States on a Quantum Computer}.
\newblock {\em IEEE Transactions on Quantum Engineering}, 1:1--17, 2020.
\newblock \href {https://doi.org/10.1109/TQE.2020.3041479}
  {\path{doi:10.1109/TQE.2020.3041479}}.

\bibitem{nielsen2010quantum}
Michael~A Nielsen and Isaac~L Chuang.
\newblock {\em Quantum computation and quantum information}.
\newblock Cambridge university press, 2010.
\newblock \href {https://doi.org/10.1017/CBO9780511976667}
  {\path{doi:10.1017/CBO9780511976667}}.

\bibitem{pallister2018optimal}
Sam Pallister, Noah Linden, and Ashley Montanaro.
\newblock {Optimal verification of entangled states with local measurements}.
\newblock {\em Physical Review Letters}, 120(17):170502, 2018.
\newblock \href {http://arxiv.org/abs/1709.03353} {\path{arXiv:1709.03353}},
  \href {https://doi.org/10.1103/PhysRevLett.120.170502}
  {\path{doi:10.1103/PhysRevLett.120.170502}}.

\bibitem{ibm-qiskit}
{Qiskit contributors}.
\newblock Qiskit: An open-source framework for quantum computing, 2023.
\newblock \href {https://doi.org/10.5281/zenodo.2573505}
  {\path{doi:10.5281/zenodo.2573505}}.

\bibitem{smolin2012efficient}
John~A. Smolin, Jay~M. Gambetta, and Graeme Smith.
\newblock {Efficient Method for Computing the Maximum-Likelihood Quantum State
  from Measurements with Additive Gaussian Noise}.
\newblock {\em Physical Review Letters}, 108(7):070502, February 2012.
\newblock \href {http://arxiv.org/abs/1106.5458} {\path{arXiv:1106.5458}},
  \href {https://doi.org/10.1103/physrevlett.108.070502}
  {\path{doi:10.1103/physrevlett.108.070502}}.

\bibitem{somma2006lower}
Rolando~D Somma, John Chiaverini, and Dana~J Berkeland.
\newblock {Lower bounds for the fidelity of entangled-state preparation}.
\newblock {\em Physical Review A}, 74(5):052302, 2006.
\newblock \href {http://arxiv.org/abs/quant-ph/0606023}
  {\path{arXiv:quant-ph/0606023}}, \href
  {https://doi.org/10.1103/PhysRevA.74.052302}
  {\path{doi:10.1103/PhysRevA.74.052302}}.

\bibitem{azure-quantinuum}
{SoniaLopezBravo,Bradben}.
\newblock {Pricing and billing plans - Azure Quantum | Microsoft Learn}, 2023.
\newblock URL:
  \url{https://learn.microsoft.com/en-us/azure/quantum/pricing?tabs=tabid-AQcredits%2Ctabid-AQcreditsQ%2Ctabid-AQcreditsRigetti%2Ctabid-learndevelop}.

\bibitem{struchalin2021experimental}
GI~Struchalin, Ya~A Zagorovskii, EV~Kovlakov, SS~Straupe, and SP~Kulik.
\newblock Experimental estimation of quantum state properties from classical
  shadows.
\newblock {\em PRX Quantum}, 2(1):010307, 2021.
\newblock \href {http://arxiv.org/abs/2008.05234} {\path{arXiv:2008.05234}},
  \href {https://doi.org/10.1103/PRXQuantum.2.010307}
  {\path{doi:10.1103/PRXQuantum.2.010307}}.

\bibitem{tokunaga2006fidelity}
Yuuki Tokunaga, Takashi Yamamoto, Masato Koashi, and Nobuyuki Imoto.
\newblock {Fidelity estimation and entanglement verification for experimentally
  produced four-qubit cluster states}.
\newblock {\em Physical Review A}, 74(2):020301, 2006.
\newblock \href {https://doi.org/10.1103/PhysRevA.74.020301}
  {\path{doi:10.1103/PhysRevA.74.020301}}.

\bibitem{wang2019optimal}
Kun Wang and Masahito Hayashi.
\newblock {Optimal verification of two-qubit pure states}.
\newblock {\em Physical Review A}, 100(3):032315, 2019.
\newblock \href {http://arxiv.org/abs/1901.09467} {\path{arXiv:1901.09467}},
  \href {https://doi.org/10.1103/PhysRevA.100.032315}
  {\path{doi:10.1103/PhysRevA.100.032315}}.

\bibitem{yu2019optimal}
Xiao-Dong Yu, Jiangwei Shang, and Otfried G{\"u}hne.
\newblock {Optimal verification of general bipartite pure states}.
\newblock {\em npj Quantum Information}, 5(1):1--5, 2019.
\newblock \href {http://arxiv.org/abs/1901.09856} {\path{arXiv:1901.09856}},
  \href {https://doi.org/10.1038/s41534-019-0226-z}
  {\path{doi:10.1038/s41534-019-0226-z}}.

\bibitem{yu2022statistical}
Xiao-Dong Yu, Jiangwei Shang, and Otfried G{\"u}hne.
\newblock Statistical methods for quantum state verification and fidelity
  estimation.
\newblock {\em Advanced Quantum Technologies}, 5(5):2100126, 2022.
\newblock \href {http://arxiv.org/abs/2109.10805} {\path{arXiv:2109.10805}},
  \href {https://doi.org/10.1002/qute.202100126}
  {\path{doi:10.1002/qute.202100126}}.

\bibitem{zhang2020experimental}
Wen-Hao Zhang, Chao Zhang, Zhe Chen, Xing-Xiang Peng, Xiao-Ye Xu, Peng Yin,
  Shang Yu, Xiang-Jun Ye, Yong-Jian Han, Jin-Shi Xu, et~al.
\newblock {Experimental optimal verification of entangled states using local
  measurements}.
\newblock {\em Physical Review Letters}, 125(3):030506, 2020.
\newblock \href {http://arxiv.org/abs/1905.12175} {\path{arXiv:1905.12175}},
  \href {https://doi.org/10.1103/PhysRevLett.125.030506}
  {\path{doi:10.1103/PhysRevLett.125.030506}}.

\bibitem{zhang2021direct}
Xiaoqian Zhang, Maolin Luo, Zhaodi Wen, Qin Feng, Shengshi Pang, Weiqi Luo, and
  Xiaoqi Zhou.
\newblock {Direct fidelity estimation of quantum states using machine
  learning}.
\newblock {\em Physical Review Letters}, 127(13):130503, 2021.
\newblock \href {http://arxiv.org/abs/2102.02369} {\path{arXiv:2102.02369}},
  \href {https://doi.org/10.1103/PhysRevLett.127.130503}
  {\path{doi:10.1103/PhysRevLett.127.130503}}.

\bibitem{zhu2019efficienthyp}
Huangjun Zhu and Masahito Hayashi.
\newblock {Efficient verification of hypergraph states}.
\newblock {\em Physical Review Applied}, 12(5):054047, 2019.
\newblock \href {http://arxiv.org/abs/1806.05565} {\path{arXiv:1806.05565}},
  \href {https://doi.org/10.1103/PhysRevApplied.12.054047}
  {\path{doi:10.1103/PhysRevApplied.12.054047}}.

\bibitem{zhu2019efficient}
Huangjun Zhu and Masahito Hayashi.
\newblock {Efficient verification of pure quantum states in the adversarial
  scenario}.
\newblock {\em {Physical Review Letters}}, 123(26):260504, 2019.
\newblock \href {http://arxiv.org/abs/1909.01900} {\path{arXiv:1909.01900}},
  \href {https://doi.org/10.1038/s41534-021-00455-6}
  {\path{doi:10.1038/s41534-021-00455-6}}.

\bibitem{zhu2019general}
Huangjun Zhu and Masahito Hayashi.
\newblock {General framework for verifying pure quantum states in the
  adversarial scenario}.
\newblock {\em Physical Review A}, 100(6):062335, 2019.
\newblock \href {http://arxiv.org/abs/1909.01943} {\path{arXiv:1909.01943}},
  \href {https://doi.org/10.1103/PhysRevA.100.062335}
  {\path{doi:10.1103/PhysRevA.100.062335}}.

\bibitem{zhu2019optimal}
Huangjun Zhu and Masahito Hayashi.
\newblock {Optimal verification and fidelity estimation of maximally entangled
  states}.
\newblock {\em Physical Review A}, 99(5):052346, 2019.
\newblock \href {http://arxiv.org/abs/1901.09772} {\path{arXiv:1901.09772}},
  \href {https://doi.org/10.1103/PhysRevA.99.052346}
  {\path{doi:10.1103/PhysRevA.99.052346}}.

\end{thebibliography}
\end{document}